\documentclass[]{aa}
\usepackage[nolist]{acronym}
\usepackage[displaymath, mathlines]{lineno}

\usepackage{graphicx}
\usepackage{txfonts}
\usepackage{natbib,twoopt}
\bibpunct{(}{)}{;}{a}{}{,} 
\usepackage{hyperref}
\usepackage{nameref}
\usepackage{cleveref}
\crefname{figure}{Fig.}{Figs.}
\Crefname{figure}{Fig.}{Figs.}
\usepackage{nccmath}
\usepackage{subcaption}
\usepackage{siunitx}
\usepackage{booktabs} 
\usepackage{multirow} 
\usepackage{threeparttable} 
\usepackage{tabularx} 
\usepackage{array} 
\usepackage{bm} 
\usepackage[table]{xcolor} 

\usepackage[hyphenbreaks]{breakurl}
\usepackage{scalerel}
\usepackage{tikz}
\usetikzlibrary{svg.path}

\definecolor{orcidlogocol}{HTML}{A6CE39}
\tikzset{
  orcidlogo/.pic={
    \fill[orcidlogocol] svg{M256,128c0,70.7-57.3,128-128,128C57.3,256,0,198.7,0,128C0,57.3,57.3,0,128,0C198.7,0,256,57.3,256,128z};
    \fill[white] svg{M86.3,186.2H70.9V79.1h15.4v48.4V186.2z}
                 svg{M108.9,79.1h41.6c39.6,0,57,28.3,57,53.6c0,27.5-21.5,53.6-56.8,53.6h-41.8V79.1z M124.3,172.4h24.5c34.9,0,42.9-26.5,42.9-39.7c0-21.5-13.7-39.7-43.7-39.7h-23.7V172.4z}
                 svg{M88.7,56.8c0,5.5-4.5,10.1-10.1,10.1c-5.6,0-10.1-4.6-10.1-10.1c0-5.6,4.5-10.1,10.1-10.1C84.2,46.7,88.7,51.3,88.7,56.8z};
  }
}

\newcommand\orcidicon[1]{\href{https://orcid.org/#1}{\mbox{\scalerel*{
\begin{tikzpicture}[yscale=-1,transform shape]
\pic{orcidlogo};
\end{tikzpicture}
}{|}}}}

\newcommand{\orcididDario}{0000-0001-9282-9462}
\newcommand{\orcididIgnas}{0000-0003-1624-3667}
\newcommand{\orcididSam}{0000-0003-4760-6168}
\newcommand{\orcididRico}{0000-0002-7261-8083}
\newcommand{\orcididYapeng}{0000-0003-0097-4414}
\newcommand{\orcididSid}{0000-0001-9552-3709}
\newcommand{\orcididAlejandro}{0000-0002-0516-7956}

\definecolor{cobalt}{rgb}{0.06, 0.2, 0.65}
\hypersetup{
  colorlinks,
  citecolor=cobalt,
  linkcolor=[rgb]{0.8, 0.2, 1.0},
  urlcolor=cobalt,
}

\definecolor{lightgray}{gray}{0.95} 

\makeatletter
\renewcommand*\aa@pageof{, page \thepage{} of \pageref*{LastPage}}

\newcommand{\logg}{\ensuremath{\log g}}

\def\kms{\ifmmode{\rm km\th s^{-1}}\else km\th s$^{-1}$\fi}
\def\th{\thinspace}
\newcommand{\Mjup}{M$_{\text{Jup}}$}

\newcommand{\twelveCO}{\textsuperscript{12}CO}
\newcommand{\thirteenCO}{\textsuperscript{13}CO}

\newcommand{\ratio}[3]{\textsuperscript{#2}#1/\textsuperscript{#3}#1}

\newcommand{\micron}{$\mu$m}
\newcommand{\vsini}{$v\sin{i}$\ }
\newcommand{\Teff}{$T_{\text{eff}}$}

\newcommand{\Cratio}{\textsuperscript{12}C/\textsuperscript{13}C}

\newcommand{\Cratiosolarlyons}{(\textsuperscript{12}C/\textsuperscript{13}C)$_{\odot}=93.5 \pm 3.1$}
\newcommand{\Cratioism}{(\textsuperscript{12}C/\textsuperscript{13}C)$_{\rm ISM}$}
\newcommand{\Cratioismilam}{(\textsuperscript{12}C/\textsuperscript{13}C)$_{\rm ISM}=68\pm15$}

\newcommand{\water}{H$_2$O}

\newcommand{\eighteenOwater}{H\textsubscript{2}\textsuperscript{18}O}
\newcommand{\CRIRES}{CRIRES$^+$}

\newcommandtwoopt{\citeads}[3][][]{\href{http://adsabs.harvard.edu/abs/#3}%
{\def\hyper@linkstart##1##2{}%
    \let\hyper@linkend\@empty\citealp[#1][#2]{#3}}}
\newcommandtwoopt{\citepads}[3][][]{\href{http://adsabs.harvard.edu/abs/#3}%
{\def\hyper@linkstart##1##2{}%
    \let\hyper@linkend\@empty\citep[#1][#2]{#3}}}

\makeatother

\begin{document} 

\title{The ESO SupJup Survey IV. Unveiling the carbon isotope ratio of GQ Lup B and its host star}

\titlerunning{Unveiling the carbon isotope ratio of GQ Lup B and its host star}
\author{D. Gonz\'{a}lez Picos       \inst{\ref{leiden}\orcidicon{\orcididDario}}
           \and I.A.G. Snellen      \inst{\ref{leiden}\orcidicon{\orcididIgnas}}
           \and S. de Regt          \inst{\ref{leiden}\orcidicon{\orcididSam}}
           \and R. Landman          \inst{\ref{leiden}\orcidicon{\orcididRico}}
           \and Y. Zhang            \inst{\ref{caltech},\ref{leiden}\orcidicon{\orcididYapeng}}
           \and S. Gandhi          \inst{\ref{warwick},\ref{warwickexoplanets},\ref{leiden}\orcidicon{\orcididSid}}
           \and A. S\'anchez-L\'opez\inst{\ref{iaa}\orcidicon{\orcididAlejandro}}
}

\newcommand{\leiden}{Leiden Observatory, Leiden University, P.O. Box 9513, 2300 RA, Leiden, The Netherlands}
\newcommand{\caltech}{Department of Astronomy, California Institute of Technology, Pasadena, CA 91125, USA}
\newcommand{\warwick}{Department of Physics, University of Warwick, Coventry CV4 7AL, UK}
\newcommand{\warwickexoplanets}{Centre for Exoplanets and Habitability, University of Warwick, Gibbet Hill Road, Coventry CV4 7AL, UK}
\newcommand{\hamburg}{Hamburger Sternwarte, Universit\"at Hamburg, Gojenbergsweg 112, 21029 Hamburg, Germany}
\newcommand{\galway}{School of Natural Sciences, Center for Astronomy, University of Galway, Galway, H91 CF50, Ireland}
\newcommand{\ipac}{IPAC, Mail Code 100-22, Caltech, 1200 E. California Boulevard, Pasadena, CA 91125, USA}
\newcommand{\mpia}{Max-Planck-Institut für Astronomie, Königstuhl 17, 69117 Heidelberg, Germany}
\newcommand{\iaa}{Instituto de Astrof{\'i}sica de Andaluc{\'i}a (IAA-CSIC), Glorieta de la Astronom{\'i}a s/n, 18008 Granada, Spain}

\institute{
    \leiden\\\email{picos@strw.leidenuniv.nl}\label{leiden}
    \and \caltech\label{caltech}
    \and \warwick\label{warwick}
    \and \warwickexoplanets\label{warwickexoplanets}
    \and \iaa\label{iaa}
}

   \date{Received 20 August 2024 / Accepted 16 December 2024}

   \abstract
   {The carbon isotope ratio (\Cratio{}) is a potential tracer of giant planet and brown dwarf formation. The GQ Lup system, which hosts the K7 T Tauri star GQ Lup A and its substellar companion GQ Lup B, provides a unique opportunity to investigate \Cratio{} in a young system.}
   {We aim to characterise GQ Lup B’s atmosphere, determining its temperature, chemical composition, spin, surface gravity, and \Cratio{}, while also measuring the \Cratio{} of its host star, GQ Lup A.}
   {We obtained high-resolution \textit{K}-band spectra of GQ Lup using CRIRES$^+$ at the VLT. We modelled GQ Lup A’s starlight contribution and fitted GQ Lup B's spectrum using atmospheric models from \texttt{petitRADTRANS}. The \Cratio{} of GQ Lup A was derived using isotope-dependent PHOENIX models.}
   {We measured atmospheric abundances in GQ Lup B, including \water{}, \twelveCO{}, \thirteenCO{}, HF, Na, Ca, and Ti, and determined a C/O ratio of $0.50 \pm 0.01$, which is consistent with the solar value. The carbon isotope ratio is \Cratio{} = $53^{+7}_{-6}$ for GQ Lup B and $51^{+10}_{-8}$ for GQ Lup A. Strong veiling of GQ Lup A’s photospheric lines was also identified and accounted for.}
   {The similar \Cratio{} values in GQ Lup A and B suggest a common origin from a shared material reservoir, supporting the idea of formation via disc fragmentation or gravitational collapse.}
 

\maketitle
%
\section{Introduction}\label{sec:introduction}

The discovery of widely separated substellar companions around young stars has challenged traditional theories of planetary formation and evolution (e.g. \citealt{chauvinGiantPlanetCompanion2005,maroisDirectImagingMultiple2008,krausLkCa15YOUNG2011}). Straddling the mass boundary between planets and brown dwarfs (BDs), these companions offer unique insights into the atmospheres of young, low-mass objects (e.g. \citealt{mayerFormationGiantPlanets2002,fortneyYoungJupitersAre2005}). Young BDs and directly imaged exoplanets share similar atmospheric properties, including high temperatures, low surface gravities, and comparable chemical compositions (e.g. \citealt{patienceSpectroscopyBrownDwarf2012,fahertyPOPULATIONPROPERTIESBROWN2016,palma-bifaniPeeringYoungPlanetary2023}). To simplify terminology, we refer to companions with masses below 75 \Mjup{} as substellar companions, classifying a lower-mass subset ($1-30$ \Mjup{}) as super-Jupiters (SJs). The current sample of SJs includes young, directly imaged companions with orbital separations of $\sim$1–300 au, effective temperatures (\Teff) of $\sim$800–2600 K, and low surface gravities (e.g. \citealt{lafreniereGeminiDeepPlanet2007,lagrangeProbableGiantPlanet2009,konopackyDetectionCarbonMonoxide2013,macintoshDiscoverySpectroscopyYoung2015,bohnYoungSunsExoplanet2020,mesaAFLepLowest2023}).

Spectroscopic studies provide critical insights into atmospheric composition, thermal structure, and dynamics, enabling the testing of formation theories (e.g. \citealt{spiegelSPECTRALPHOTOMETRICDIAGNOSTICS2012}). Substellar companions are thought to form via two main pathways: core accretion \citep{pollackFormationGiantPlanets1996,johansenExploringConditionsForming2019} and disc fragmentation or gravitational instability \citep{bossGiantPlanetFormation1997,mayerFormationGiantPlanets2002}. In core accretion, solid cores form first, followed by the accretion of gas from the protoplanetary disc. While this process explains the formation of Solar System gas giants, it struggles to account for planets at separations greater than 30 au \citep{inabaFormationGasGiant2003,alibertModelsGiantPlanet2005}. Gravitational instability, on the other hand, involves the direct collapse of disc material and can produce more massive objects at distances of 20–60 au \citep{bossFORMATIONGIANTPLANETS2011,vorobyovFormationGiantPlanets2013}. However, this mechanism fails to fully explain the population of substellar companions at 100–200 au \citep{dodson-robinsonFORMATIONMECHANISMGAS2009}. Subsequent dynamical interactions, such as disc-planet or planet-planet scattering, migration, or ejection, likely influence the final positions of these objects \citep{verasFormationSurvivalDetectability2009,boleyINTERACTIONSMODERATELONGPERIOD2012,bitschFormationPlanetarySystems2019,carterSurvivabilityPopulationGas2023}.

The detection and characterisation of widely separated substellar companions rely on high-contrast imaging instruments capable of spatially resolving faint objects next to bright stars (e.g. \citealt{macintoshFirstLightGemini2014,beuzitSPHEREExoplanetImager2019}). Multi-wavelength photometric and spectroscopic observations reveal physical properties such as effective temperature, surface gravity, and chemical composition (e.g. \citealt{barmanCLOUDSCHEMISTRYATMOSPHERE2011,konopackyDetectionCarbonMonoxide2013,hoeijmakersMediumresolutionIntegralfieldSpectroscopy2018}). High-resolution spectroscopy (HRS) has revolutionised atmospheric studies, enabling precise retrievals of chemical composition, thermal structure, and cloud properties, as well as measurements of spin and surface gravity (e.g. \citealt{snellenFastSpinYoung2014,ruffioDeepExplorationPlanets2021,whitefordRetrievalStudyCool2023,landmanPictorisEyesUpgraded2024}).

The carbon isotopic ratio (\Cratio) has been proposed as a tracer of formation pathways for SJs and BDs. Like the C/O ratio, the \Cratio{} is possibly sensitive to the formation environment, objects formed via accretion of solid material are expected to inherit the \Cratio{} of the solid material, while objects formed via gravitational instability are expected to have a \Cratio{} similar to the local interstellar medium \citep{obergAstrochemistryCompositionsPlanetary2021, zhang12CO132021}. Isotope fractionation processes (e.g. \citealt{langer12C13CIsotope1993,visserPhotodissociationChemistryCO2009}) in molecular clouds and protoplanetary discs can alter the CO isotopologue ratios, leading to variations in the \Cratio{} across different regions of the disc \citep{yoshidaNewMethodDirect2022,leeCarbonIsotopeChemistry2024}. The first measurement of the \Cratio{} in a substellar companion was reported for YSES 1 b \Cratio{} = $30^{+10}_{-7}$ \citep{zhang13COrichAtmosphereYoung2021}, employing an infrared medium-resolution spectrograph. The \Cratio{} of YSES 1 b was found to be significantly lower than that of the local ISM (\Cratioismilam; \citealt{milam1213Isotope2005}) and clearly lower than the Sun's carbon isotope ratio (\Cratiosolarlyons; \citealt{lyonsLightCarbonIsotope2018}), suggesting that the object was enriched in ${}^{13}$C via the accretion of ${}^{13}$C-rich ices beyond the snow line. A subsequent study of the atmosphere of the young isolated BD 2M 0355 found a \Cratio{} of $97^{+25}_{-18}$, a significantly higher value than that of YSES 1 b, suggesting that the two objects formed via different mechanisms \citep{zhang12CO132021}. Additional data from the science verification of the upgraded CRyogenic high-resolution InfraRed Echelle Spectrograph (CRIRES$^+$; \citealt{dornCRIRESSkyESO2023}) confirmed the detection of the \Cratio{} with a value of $108 \pm 10$ and hinted at the detection of C${}^{18}$O in the atmosphere of 2M 0355 \citep{zhangVLTCRIRESScience2022}. Recent work with the Keck Planet Imager and Characterizer (KPIC; \citealt{delormeKeckPlanetImager2020}) also measured the \Cratio{} in the HIP 55507 system, revealing values of $98^{+28}_{-22}$ and $79^{+21}_{-16}$ for the companion HIP 55507 B and the host star HIP 55507 A, respectively \citep{xuanValidationElementalIsotopic2024}, suggesting a homogeneous isotopic composition. \citealt{costesFreshViewHot2024} also analysed the KPIC spectrum of HD 984 B and determined a C/O=$0.50 \pm 0.1$ and a \Cratio{} of $98^{+20}_{-25}$, suggesting that the object formed via gravitational collapse or disc instability. Measurements from space-based instruments, namely JWST, have also been used to measure the \Cratio{} in the atmosphere of the young SJ VHS 1256 b, revealing a value of $62 \pm 2$ and a super-solar C/O ratio of $0.69 \pm 0.01$, in addition to placing constraints on \ratio{O}{16}{18} and \ratio{O}{16}{17} that indicate values lower than that of the local ISM for the oxygen isotope ratios yet a \Cratio{} consistent with the lower value of the ISM \citep{gandhiJWSTMeasurements132023}.

To expand the sample of SJs and BDs with measured \Cratio{} values and assess the role of isotope ratios as formation tracers, the ESO SupJup Survey (PI: Snellen; \citealt{deregtESOSupJupSurvey2024}) aims to characterise the atmospheres of young substellar companions and isolated BDs using HRS with VLT/\CRIRES{}. The initial results of this survey were reported by \citealt{deregtESOSupJupSurvey2024}, who analysed the atmosphere of a field L-dwarf was analysed and found a \Cratio{} of $184^{+61}_{-40}$. Further characterisation of the \Cratio{} ratios in three young BDs with similar spectral types and ages as GQ Lup B was presented by \citealt{gonzalezpicosESOSupJupSurvey2024} who found values of $81^{+28}_{-19}$, $79^{+20}_{-14}$, and $114^{+69}_{-33}$ for TWA 28, 2M J0856, and 2M J1200, respectively. In the third paper of the series, \citealt{zhangESOSupJupSurvey2024} confirmed the presence of \thirteenCO{} in the atmosphere of YSES 1 b, refining the \Cratio{} to $81\pm9$ and reporting the detection of atmospheric CO and \water{} in YSES 1 c.

In this work we analyse the atmosphere of GQ Lup B as part of the ongoing ESO SupJup survey, focusing on its temperature profile, chemical composition, spin, surface gravity, and the determination of the \Cratio{} in its atmosphere and that of its host star, GQ Lup A. The paper is structured as follows: Sect. \ref{sec:gqlup} provides an overview of the GQ Lup system; Sect. \ref{sec:observations} details the observations and data reduction; Sect. \ref{sec:atmospheric_retrieval} presents the atmospheric retrieval of GQ Lup B; Sect. \ref{sec:modelling_gqlup_a} describes the modelling of the spectrum of the host star; Sect. \ref{sec:results} reports the results; Sect. \ref{sec:discussion} discusses the findings and compares them with previous studies; and Sect. \ref{sec:conclusions} presents the conclusions.

\section{The GQ Lup system}\label{sec:gqlup}

The GQ Lup system, situated in the Lupus I cloud \citep{tachihara13COObservationsLupus1996}, is a young stellar system comprising GQ Lup A, a K7Ve-type T Tauri star enveloped by a circumstellar disc \citep{daiMILLIMETERDUSTEMISSION2010,macgregorALMAMEASUREMENTSCIRCUMS2017}, and a substellar companion, GQ Lup B, located at a wide angular separation of approximately 0.7", equivalent to $\sim 140$ au \citep{neuhauserEvidenceComovingSubstellar2005}. The properties of the system are listed in \Cref{tab:prop_gqlup}. Initial mass estimates for GQ Lup B ranged from 10 to 40 \Mjup{} \citep{maroisGQLupVisible2006,mcelwainFirstHighContrastScience2007}, covering a spectrum from planetary-mass objects to BDs; however, to date, its mass remains poorly constrained. Near-infrared spectroscopy has established GQ Lup B as a late M- to early L-type object, with \Teff $\approx$ 2100-2700 K and \logg $\leq 4.5$ \citep{neuhauserEvidenceComovingSubstellar2005,maroisGQLupVisible2006,mcelwainFirstHighContrastScience2007,seifahrtNearinfraredIntegralfieldSpectroscopy2007,lavigneNEARINFRAREDOBSERVATIONSGQ2009,stolkerCharacterizingProtolunarDisk2021}. The system's estimated age, between 2 and 5 Myr \citep{donatiMagnetometryClassicalTauri2012}, aligns with the observed low-surface gravity features \citep{seifahrtNearinfraredIntegralfieldSpectroscopy2007}. Through HRS cross-correlation techniques, \citealt{schwarzSlowSpinYoung2016} identified CO and H2O within GQ Lup B's atmosphere and determined its rotational velocity to be $v\sin{i} = 5.3^{+1.0}_{-0.9}$ \kms, a slow spin characteristic of its young age. Studies focusing on the orbital dynamics of the GQ Lup system have estimated the inclination of GQ Lup B's orbit to be $i \approx 60$ deg and the semi-major axis to be $a = 100-150$ au \citep{jansonEarlyComeOnAdaptive2006,ginskiAstrometricFollowupObservations2014,schwarzSlowSpinYoung2016}.

Further analysis combining optical, near-, and mid-infrared data confirmed GQ Lup B's spectral type as M9 and highlighted low surface gravity features \citep{stolkerCharacterizingProtolunarDisk2021}. Additionally, the presence of mid-infrared excess in GQ Lup B's spectral energy distribution and accretion signatures in the optical spectrum suggested the existence of a proto-lunar disc \citep{stolkerCharacterizingProtolunarDisk2021}, positioning this system as a prime target for investigations into satellite formation around substellar companions \citep{lazzoniDetectabilitySatellitesDirectly2022}. Recent JWST/MIRI \citep{gardnerJamesWebbSpace2023,argyriouJWSTMIRIFlight2023} of the GQ Lup system in the mid-infrared have confirmed the presence of a disc around GQ Lup B, offering insights into the disc's extent and temperature \citep{cugnoMidInfraredSpectrumDisk2024}.

A newly released analysis of the \textit{K}-band spectrum of GQ Lup B with KPIC has reported the detection of \thirteenCO{} in its atmosphere \citep{xuanAreThesePlanets2024}. The authors of this study also provide estimates of the elemental composition and the carbon isotope ratio of a small sample of substellar companions with masses ranging from 10 to 30 \Mjup. \citealt{xuanAreThesePlanets2024} report a \Cratio{} of approximately $150$ and a super-solar C/O of approximately $0.70$ for GQ Lup B (see Sect. \ref{sec:discussion} for a comparison with our results).

A recent addition to the GQ Lup system is the discovery of 2MASS J15491331-3539118, a low-mass star bound to the GQ Lup system, designated GQ Lup C \citep{alcala2MASSJ154913313539118New2020}. Positioned at a projected separation of 16.0", approximately 2400 au away, GQ Lup C adds to the architectural diversity of the GQ Lup system. Its near-infrared spectrum indicates M4 spectral type, with effective temperature around 3200 K, mass of approximately 157 \Mjup{}, and a notably low surface gravity of $3.74 \pm 0.24$ \citep{alcala2MASSJ154913313539118New2020}. The identification of GQ Lup C not only adds valuable context to the formation pathways within the GQ Lup system but also contributes to broader discussions on the formation of companions with extremely wide orbits around young stellar objects.

\renewcommand{\arraystretch}{1.3}

\begin{table}
    \caption{Properties of GQ Lup A and B.}
    \label{tab:prop_gqlup}
    \begin{tabular}{@{}lcl@{}}
    \toprule
    Parameter & Value & References \\
    \midrule
    \multicolumn{3}{c}{GQ Lup A} \\
    \midrule
    Spectral Type & K7Ve & 1 \\
    $A_V$ (mag) & $0.4\pm0.2$ & 2 \\
    log($L/L_\odot$) & $0.0\pm0.1$ & 3, 4 \\
    $T_{\mathrm{eff}}$ (K) & $4300\pm50$ & 3, 4 \\
    Radius & $1.7\pm0.2~R_\odot$ & 3 \\
    Mass & $1.05\pm0.07~M_\odot$ & 3,4,5,6 \\
    log $\dot{M}$ ($M_\odot~\mathrm{yr}^{-1}$) & $-9$ to $-7$ & 3, 7 \\
    log $g$ & $3.7\pm0.2$ & 3, 4 \\
    Inclination (\degr) & $27\pm5$ & 8 \\
    $v~\mathrm{sin}(i)$ (km $\mathrm{s}^{-1}$) & $5\pm1$ & 3, 9 \\
     & $5.37\pm0.05$ & This work \\
    Rotation Period (d) & $8.45\pm0.20$ & 8 \\
    \midrule
    \multicolumn{3}{c}{GQ Lup B} \\
    \midrule
    Separation (\arcsec) & $0.721\pm0.003$ & 7 \\
    PA (\degr) & $277.6\pm 0.4$ & 7 \\
    Spectral Type & M9 & 10 \\
    $A_V$ (mag) & $\approx$ 2.3 & 10, 11, 14 \\
    log($L/L_\odot$) & $-2.47\pm0.28$ & 3, 4 \\
    $T_{\mathrm{eff}}$ (K) & $2300-2700$ & 3, 4, 10, 14 \\
    Radius & $3.5-4.5$ $R_{\mathrm{Jup}}$ & 3, 10, 11, 14 \\
    Mass & $\sim10$--$40~M_{\mathrm{Jup}}$ & 3, 4, 5, 6, 12, 13 \\
    log $g$ & $4.0\pm0.5$ & 3, 4, 10   \\
     & $3.83^{+0.17}_{-0.18}$ & This work \\
    $v~\mathrm{sin}(i)$ (km $\mathrm{s}^{-1}$) & $5.3^{+0.9}_{-1.0}$ & 4, 9 \\
     & $5.56 \pm 0.02$ & This work \\
    \bottomrule
    \end{tabular}
    
\caption*{References:
    (1) \citealt{herbigPropertiesProblemsTauri1962}, 
    (2) \citealt{batalhaVariabilitySouthernTauri2001}, 
    (3) \citealt{donatiMagnetometryClassicalTauri2012}, 
    (4) \citealt{lavigneNEARINFRAREDOBSERVATIONSGQ2009}, 
    (5) \citealt{neuhaeuserAstrometricPhotometricMonitoring2008},
    (6) \citealt{maroisGQLupVisible2006}, 
    (7) \citealt{wuALMAMagAOStudy2017},
    (8) \citealt{broegRotationalPeriodGQ2007},
    (9) \citealt{schwarzSlowSpinYoung2016}, 
    (10) \citealt{stolkerCharacterizingProtolunarDisk2021}, 
    (11) \citealt{cugnoMidInfraredSpectrumDisk2024}, 
    (12) \citealt{mcelwainFirstHighContrastScience2007}, 
    (13) \citealt{seifahrtNearinfraredIntegralfieldSpectroscopy2007}, 
    (14) \citealt{demarsEmissionLineVariability2023}.
    }
\end{table}

\section{Observations}\label{sec:observations}

We obtained high-resolution spectra of the GQ Lup system on February 28, 2023, as part of the ESO SupJup survey, using the \CRIRES{} instrument at the VLT \citep{dornCRIRESSkyESO2023}. \CRIRES{} is an upgraded version of CRIRES, featuring a long-slit, cross-dispersed echelle spectrograph equipped with the MACAO adaptive optics system \citep{paufiqueMACAOCRIRESStepHighresolution2004}.

\begin{figure}
    \centering
    \includegraphics[width=\columnwidth]{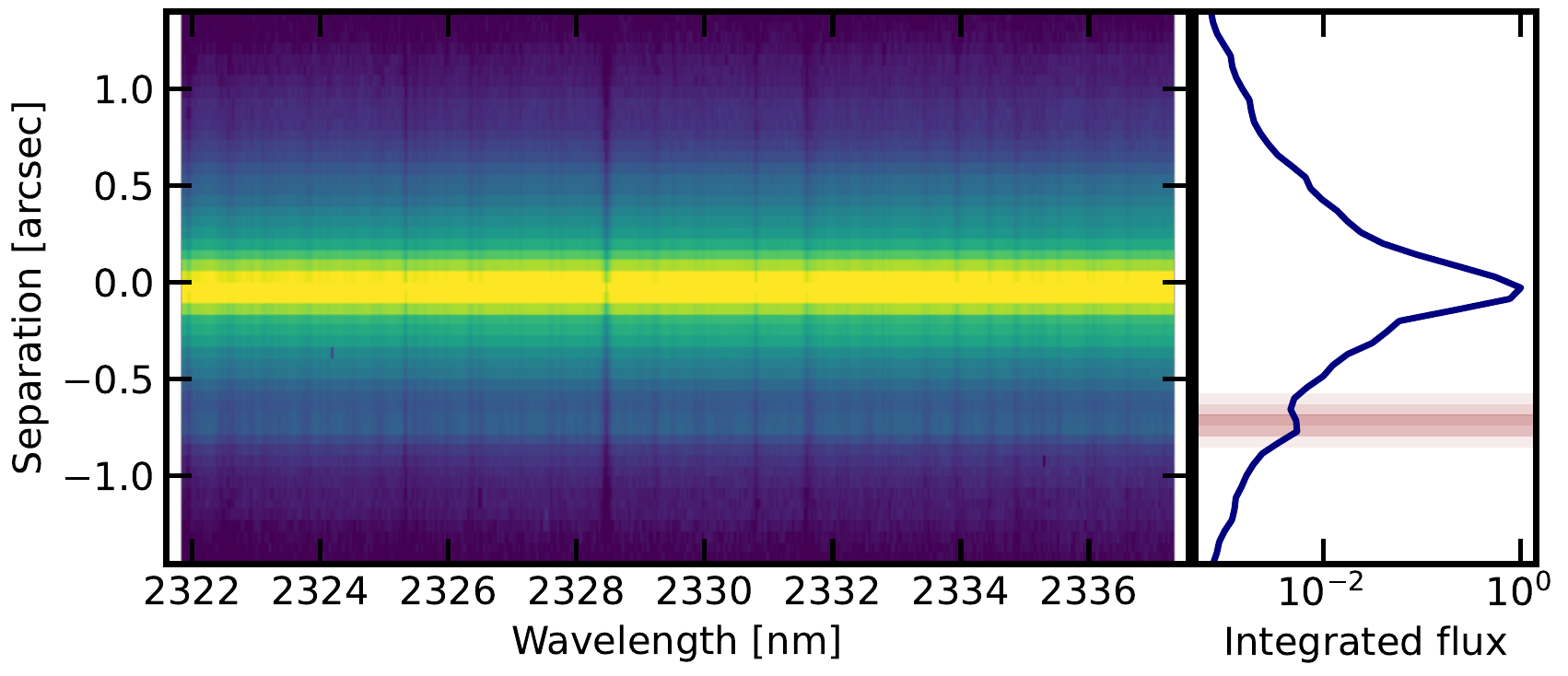}
    \caption{Calibrated 2D image of a single order-detector pair. The vertical axis is the spatial direction and the horizontal axis the spectral direction. The bright central source is GQ Lup A, and the fainter source at -0.71 arcseconds is GQ Lup B. The telluric lines are visible as vertical stripes in the image. The right panel shows the integrated line spread function and the spatial profile employed for the extraction of the data at the position of GQ Lup B.}
    \label{fig:fig1_2d_spectrum}
\end{figure}

Observations were conducted in the 0.2" narrow-slit mode, achieving a resolving power of $\mathcal{R} \sim 120,000$, as determined from telluric line fits. This value exceeds the nominal $\mathcal{R} = 100,000$ under favourable observing conditions (see \cref{app:obs_conditions}), consistent with prior findings by \citealt{holmbergFirstLookCRIRES2022}. The slit was oriented at a position angle of 278.91 degrees \citep{ginskiAstrometricFollowupObservations2014}, enabling simultaneous observation of GQ Lup A and GQ Lup B (see \Cref{fig:fig1_2d_spectrum}). The K2166 wavelength setting was used, covering the \textit{K}-band spectral range from 1.92 to 2.48 \micron, with particular focus on the \thirteenCO{} band heads near 2.345 \micron. The total exposure time was 3 hours, split into 60 exposures of 180 seconds each, using an ABBA nodding sequence with a 5" nod throw.

Sky conditions were photometric and stable, with an average seeing of 0.6" and an airmass ranging from 1.5 to 1.0 (see Appendix \ref{app:obs_conditions}). To aid in removing Earth's atmospheric transmission features, the standard star HD 1407 was observed immediately before GQ Lup using the same slit width and nodding sequence, though at a slightly higher airmass. To account for this discrepancy, we applied forward-modelling of the telluric transmission rather than directly removing it from the data (see \Cref{subsec:telluric_transmission}).

Data reduction was performed using the Python package \texttt{excalibuhr}\footnote{\url{https://github.com/yapenzhang/excalibuhr}} \citep{zhangESOSupJupSurvey2024}. The reduction pipeline included the following steps:

\begin{enumerate}
    \item \emph{Flat-fielding}: Correcting for pixel-to-pixel variations in the detector using flat-field frames.
    \item \emph{Spectral order tracing}: Identifying the traces of the spectral orders by fitting a second-order polynomial to the spatial profile from the combined flat-field frame.
    \item \emph{Spectral order curvature tracing}: Determining the curvature of the spectral orders from Fabry-Pérot calibration frames and using it as an initial wavelength solution.
    \item \emph{Blaze extraction}: Extracting the blaze function from the flat-field frames and using it to correct the spectral orders.
    \item \emph{Nodding pair subtraction}: Subtracting the nodding pairs to remove the sky background.
    \item \emph{Nodding pair combination}: Generating a combined image for each nodding position by taking the median value of all frames in the same nodding position.
    \item \emph{Spectral extraction}: Extracting the spectra using optimal extraction \citep{horneOPTIMALEXTRACTIONALGORITHM1986} with a Gaussian spatial profile. The extraction is performed at two different slit positions: one centred on GQ Lup A and the other on GQ Lup B. The size of the half-aperture for the extraction is set to 8 and 2 pixels for GQ Lup A and GQ Lup B, respectively, corresponding to 0.45" and 0.11" on the sky.
    \item \emph{Wavelength solution calibration}: Calibrating the wavelength solution using telluric lines present in the spectra and a telluric model matching the observing conditions obtained from \texttt{Skycalc}\footnote{\url{http://www.eso.org/observing/etc/skycalc/}}. The telluric model is convolved with a Gaussian kernel to match the instrumental resolution. The best-fit wavelength solution is obtained by maximising the cross-correlation between the observed and telluric models shifted in wavelength by a second-order polynomial.
\end{enumerate}

\section{Atmospheric retrieval of GQ Lup B}\label{sec:atmospheric_retrieval}

\subsection{Linear model}\label{subsec:linear_model}
The spectrum at the position of GQ Lup B was modelled using a linear framework, which assumes that the observed signal is a combination of the atmospheric model of the companion and the scattered starlight from GQ Lup A \citep{hochModerateresolutionKbandSpectroscopy2020,ruffioDeepExplorationPlanets2021,xuanValidationElementalIsotopic2024}. The model is expressed as:

\begin{align}
  \mathbf{d} &= \mathbf{\phi} \mathbf{M} + \mathbf{n} = \underbrace{\mathbf{\phi}_{\text{0}} \mathbf{M}_{\text{0}}}_{\text{companion}} + \underbrace{\mathbf{\phi}_{\text{1...i}} \mathbf{M}_{\text{1...i}}}_{\text{starlight}} + \mathbf{n},
\end{align}

where $\mathbf{d}$ is the extracted spectrum at the position of GQ Lup B, $\mathbf{M}$ is the model matrix, $\mathbf{\phi}$ is the vector of linear coefficients, and $\mathbf{n}$ is the noise. The coefficient $\mathbf{\phi}_{\text{0}}$ represents the amplitude of the companion model, while $\mathbf{\phi}_{\text{1...i}}$ represent the amplitudes of the starlight model components. 

The matrix $\mathbf{M}_0$ contains the atmospheric model of the companion, generated using \texttt{petitRADTRANS} based on a set of non-linear parameters $\psi$ (see \Cref{subsec:model_gqlup_b}). The matrix $\mathbf{M}_{1...i}$ represents the scattered starlight from GQ Lup A, derived from its observed spectrum (see \Cref{app:starlight}). The linear coefficients are determined at each likelihood evaluation using the procedure outlined in Eq. \ref{eq:nnls}.

The likelihood function follows \cite{ruffioRadialVelocityMeasurements2019} and is expressed as:

\begin{equation}
    \begin{aligned}
    \mathcal{L}(\psi, \phi, s^2) &= \frac{1}{\sqrt{(2\pi)^{N_{d}}|\Sigma_0| s^{2N_{d}}}} \\
    &\quad\cdot \exp\left(-\frac{1}{2s^2}\chi^2_{0,\phi=\tilde{\phi},\psi}\right) \\
    &\quad\cdot \exp\left(-\frac{1}{2s^2}\Delta{\phi}^{\top}(M^{\top}\Sigma_0^{-1}M)\Delta{\phi}\right),
    \end{aligned}
    \label{eq:likelihood}
\end{equation}

where $N_{d}$ is the number of data points, $\Sigma_0$ is the covariance matrix, $s^2$ is the noise scaling factor, $\chi^2_{0,\phi=\tilde{\phi},\psi}$ is the optimal chi-squared value, and $\Delta{\phi}$ represents the uncertainty in the linear parameters, $\phi = \tilde{\phi} + \Delta{\phi}$. The optimal linear parameters $\tilde{\phi}$ and noise scaling factors $\tilde{s^2}$ are obtained by minimising the chi-squared value:

\begin{align}
    \begin{aligned}
      \chi_0^2 &= \left(\mathbf{d} - \mathbf{\phi} \mathbf{M}\right)^{\top} \Sigma_0^{-1} \left(\mathbf{d} - \mathbf{\phi} \mathbf{M}\right), \\
      \nabla_{\phi}\chi_0^2 &= 0 \implies \tilde{\phi} = \left(\mathbf{M}^{\top} \Sigma_0^{-1} \mathbf{M}\right)^{-1} \mathbf{M}^{\top} \Sigma_0^{-1} \mathbf{d},\\
      \nabla_{s^2}\chi_0^2 &= 0 \implies \tilde{s^2} = \frac{1}{N_{d}}\left(\mathbf{d} - \mathbf{\tilde{\phi}} \mathbf{M}\right)^{\top} \Sigma_0^{-1} \left(\mathbf{d} - \mathbf{\tilde{\phi}} \mathbf{M}\right) = \frac{\chi^2_{0,\phi=\tilde{\phi}}}{N_{d}},
    \end{aligned}
\label{eq:nnls}
\end{align}

 which in practise is solved using the non-negative least squares algorithm \citep{lawsonSolvingLeastSquares1995}\footnote{\href{https://docs.scipy.org/doc/scipy/reference/generated/scipy.optimize.nnls.html}{scipy.optimize.nnls}}. To marginalise over the linear parameters and noise scaling factor, we integrated the likelihood function across the parameter space (see Appendix D of \citealt{ruffioRadialVelocityMeasurements2019}), which yields the log-likelihood function

\begin{align}
  \log{\mathcal{L}(\psi)} &= -\frac{N_d-N_{\phi}}{2} \log{2\pi}
    - \frac{1}{2} \log{|{\Sigma_0}|} \notag\\&\quad  + \log{\Gamma\left(\frac{N_d-N_{\phi}+1}{2}\right)} - \frac{1}{2} \log{|{M^\top \Sigma_0^{-1} M}|}\notag \\
    &\quad- \frac{N_d-N_{\phi}+1}{2}\log{\chi^2_{0,\phi=\tilde{\phi},\psi}} 
    , \label{eq:log_likelihood}
\end{align}

where $N_{\phi}$ is the number of linear parameters, and $\log \Gamma(n)$ denotes the logarithm of the gamma function\footnote{\href{https://docs.scipy.org/doc/scipy/reference/generated/scipy.special.loggamma.html}{scipy.special.loggamma}}.

\subsection{Correlated noise}\label{subsec:correlated_noise}
Data uncertainties often exhibit correlations, potentially biasing parameter estimates and underestimating uncertainties \citep{kawaharaAutodifferentiableSpectrumModel2022}. To account for correlated noise, we introduced a noise model that incorporates a covariance matrix $\Sigma_0$ in the likelihood function. The covariance matrix is defined following \cite{deregtESOSupJupSurvey2024} and \cite{gonzalezpicosESOSupJupSurvey2024}:

\begin{equation}
  \Sigma_{0,ij} = \delta_{ij} \sigma_{i}^2 + k_{ij},
\end{equation}

where $i$ and $j$ index data points, $\delta_{ij}$ is the Kronecker delta, $\sigma_{i}$ is the uncertainty of data point $i$, and $k_{ij}$ represents the Gaussian process (GP) kernel. The GP kernel is given by:

\begin{equation}
  k_{ij} = a^2\left(\frac{\sigma_{i} + \sigma_{j}}{2}\right)^2 \exp\left(-\frac{(x_i - x_j)^2}{2l^2}\right),
\end{equation}

where $a$ and $l$ are free parameters representing the kernel amplitude and length scale, respectively.

\subsection{Starlight model}

\begin{figure*}[ht!]
  \centering
  \includegraphics[width=\textwidth]{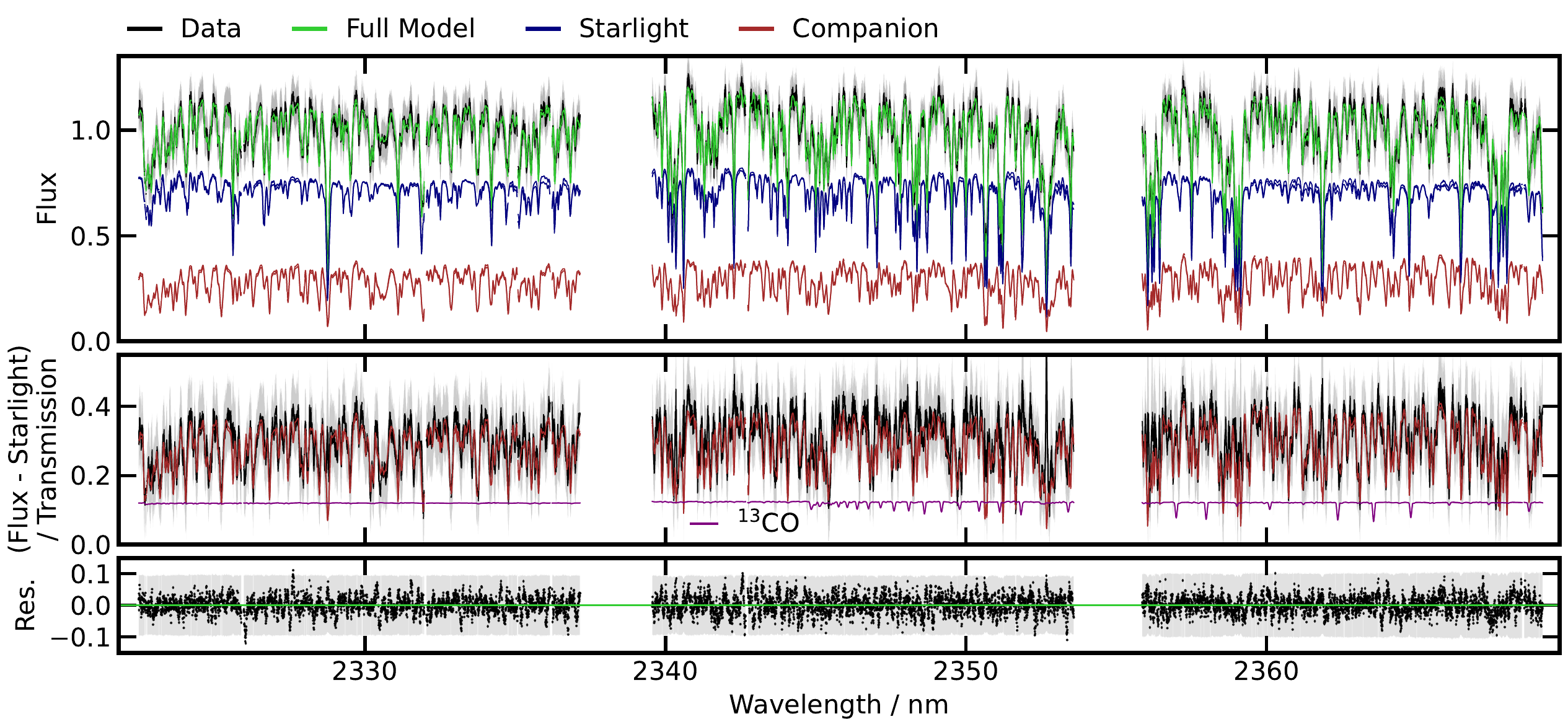}
  \caption{Best-fit model of the spectrum at the position of GQ Lup B. Top: Observed spectrum (black), joint best-fit model (green) and components of the starlight model and companion (blue and brown, respectively). Centre: Data with the best-fit starlight model subtracted compared to the companion model. Bottom: Residuals of the best-fit model. The shaded region represents the root-squared diagonal of the covariance matrix with the optimal noise scaling factor.}
  \label{fig:fig2_bestfit}
\end{figure*}

The signal at the position of the companion is dominated by starlight, as illustrated in \Cref{fig:fig2_bestfit}. The on-axis starlight is scattered across the slit due to atmospheric effects and telescope optics \citep{viganExoplanetCharacterizationLong2008}. To model the scattered starlight at the position of the companion, we employed a technique involving a linear combination of different versions of the observed on-axis spectrum of GQ Lup A, adjusted with different velocity shifts and scaling factors \citep{landmanPictorisEyesUpgraded2024}.

 Our approach involves constructing a model for the starlight using shifted versions of the observed on-axis spectrum of GQ Lup A. We considered a range of pixel shifts denoted by $N_s=2s+1$, where $s$ ranges from -3 to 3 in steps of 1 pixel, equivalent to a maximum shift of 9 km/s, which sufficiently covers the observed broadening of the spectrum due to scattering (see \cref{app:starlight} for more details).

Moreover, we accounted for low-frequency variations in the starlight model, which may stem from various sources such as atmospheric conditions, telescope optics, and detector characteristics. To model these variations, we adopted the spline decomposition method introduced in \citep{ruffioDetectingExomoonsRadial2023}. With each shifted component of the model, we generated a cubic spline decomposition with $N_k=10$ knots, corresponding to approximately 200 pixels in width. The complete starlight model is represented as follows:

\begin{equation}
\mathbf{M}_{\text{starlight}} = [\mathbf{m}_{s=-3, k=0} \ldots \mathbf{m}_{s=3, k=9}],
\end{equation}

 where the linear coefficients of the starlight model components are independent for each order-detector ($N_{\text{od}}=21$) and nodding position ($N_{\text{ab}}=2$) and are optimised at every evaluation of the log-likelihood function (see Eq. \ref{eq:log_likelihood}). The total number of linear parameters for a single order-detector pair of a single nodding position is $N_{\phi} = N_{\text{s}} \times N_{\text{k}} = 90$, and the total number of linear parameters for all order-detector pairs and nodding positions is $N_{\phi} = N_{\text{od}} \times N_{\text{ab}} \times N_{\text{s}} \times N_{\text{k}} = 3780$.

\subsection{Atmospheric model}\label{subsec:model_gqlup_b}

To accurately characterise the high-resolution spectra of GQ Lup B, we employed the radiative transfer code \texttt{petitRADTRANS}\footnote{\url{https://petitradtrans.readthedocs.io/en/2.7.7/}} (v2.7; \citealt{mollierePetitRADTRANSPythonRadiative2019}). The outgoing spectrum is calculated by solving the radiative transfer equation within a 1D plane-parallel atmosphere, where each layer is characterised by its temperature, pressure, and mass fractions of relevant chemical species. The atmospheric model is defined by several key parameters, including the temperature profile, chemical composition, and surface gravity of the object.

\subsubsection{Chemical composition}\label{subsec:chemical_composition}
The atmospheres of substellar objects, such as GQ Lup B, exhibit rich molecular compositions, that prominently feature water (\water{}) and carbon monoxide (CO). Our atmospheric model incorporates a variety of line species, continuum, and collision-induced absorption (CIA) to accurately reproduce the spectral characteristics observed in GQ Lup B (see \Cref{tab:opacity_references} for a comprehensive list of species and their corresponding references).
\begin{table}[h!]
    \centering
    \caption{Opacity species and references}
    \begin{tabular}{cc}
    \hline
    Species & References \\ \hline
    \multicolumn{2}{c}{Line Species} \\ \hline
    $^{12}$CO, $^{13}$CO, C$^{18}$O & \cite{rothmanHITEMPHightemperatureMolecular2010}, \cite{liROVIBRATIONALLINELISTS2015} \\
    H$_2$$^{16}$O & \cite{polyanskyExoMolMolecularLine2018} \\
    H$_2$$^{18}$O & \cite{polyanskyExoMolMolecularLine2017} \\
    HF & \cite{wilzewskiH2HeCO22016} \\
    Na, K & \cite{allardNewStudyLine2019a} \\
    Mg, Ca, Ti & \cite{castelliNewGridsATLAS92003} \\
    \hline
    \multicolumn{2}{c}{Continuum Species} \\ \hline
    H$_2$-H$_2$ & \cite{dalgarnoRayleighScatteringMolecular1962} \\
    H$_2$-He & \cite{chanRefractiveIndexHelium1965} \\
    H$^-$ & \cite{grayObservationAnalysisStellar2022} \\\hline
    \multicolumn{2}{c}{CIA Species} \\ \hline
    H$_2$-H$_2$ & \cite{borysowCollisioninducedRototranslationalAbsorption1988} \\\hline
    \end{tabular}
    \label{tab:opacity_references}
\end{table}

We adopted a free composition approach, treating the volume mixing ratio (VMR) of each species as a free parameter instead of relying on chemical equilibrium calculations. This method provides greater flexibility, allowing us to explore a broader range of chemical compositions, which can result in better model fits \citep{gandhiJWSTMeasurements132023, deregtESOSupJupSurvey2024,gonzalezpicosESOSupJupSurvey2024}.

From the retrieved abundances, expressed as volume mixing ratios $n_X$, we can infer relevant atmospheric properties and derive elemental and isotopic ratios of interest. For instance, the C/O ratio is derived from the abundances of key carbon- and oxygen-bearing species:

\begin{align}\label{eq:C_O_ratio}
  \mathrm{C/O} = \frac{n_{\mathrm{\twelveCO}}+n_{\mathrm{\thirteenCO}}}{n_{\mathrm{H_2O}}+n_{\mathrm{\twelveCO}}+n_{\mathrm{\thirteenCO}}}.
\end{align}

Similarly, the ratio of \twelveCO/\thirteenCO{} provides insights into the isotopic composition of carbon:

\begin{align}\label{eq:12CO_13CO_ratio}
\mathrm{^{12}C/^{13}C} = \frac{n_{\mathrm{\twelveCO}}}{n_{\mathrm{\thirteenCO}}}.
\end{align}

Furthermore, we can estimate the atmospheric metallicity using the carbon-to-hydrogen ratio normalised to solar metallicity:

\begin{align}\label{eq:metallicity}
\mathrm{[C/H]} = \log_{10}\left(\frac{n_{\mathrm{C}}}{n_{\mathrm{H}}}\right) - \log_{10}\left(\frac{n_{\mathrm{C}}}{n_{\mathrm{H}}}\right)_{\odot},
\end{align}

where the solar value is $\log_{10}\left(\frac{n_{\mathrm{C}}}{n_{\mathrm{H}}}\right)_{\odot} = -3.54$ \citep{asplundChemicalMakeupSun2021}.

\subsubsection{Temperature profile}\label{subsec:temperature_profile}
For widely separated companions like GQ Lup B, which receive minimal irradiation from their host stars, the internal heat flux primarily dictates the temperature profile. As one moves from high to low pressures within the atmosphere, the temperature typically decreases with altitude, following a gradient, $\frac{d\log{T}}{d\log{P}} = \nabla$, that reflects the dominant energy transfer mechanism.

In the denser, deeper regions of the atmosphere, energy transport is primarily governed by convective processes, which ideally follow the adiabatic temperature gradient ($\nabla = \nabla_{\text{ad}}$; \citealt{baraffeEvolutionaryModelsLowmass2002}). As the altitude increases and the atmosphere becomes less dense, the mean free path of photons grows, causing some wavelength regions to become optically thin. This shift allows radiation to emerge as the dominant mechanism of energy transfer. The transition between these energy transport regimes defines the radiative-convective equilibrium (RCE; \citealt{marleyCoolSideModeling2015}) boundary.

Various parameterisations for temperature profiles have been proposed, ranging from physically motivated models to more flexible, free-form models. Physically motivated models attempt to describe the temperature structure using parameters with direct physical interpretations \citep{madhusudhanTEMPERATUREABUNDANCERETRIEVAL2009,molliereRetrievingScatteringClouds2020a,lineINFORMATIONCONTENTEXOPLANETARY2012}. On the other hand, traditional planetary science models often retrieve temperature at each discrete atmospheric layer \citep{rodgersInverseMethodsAtmospheric2000,irwinNEMESISPlanetaryAtmosphere2008a}. However, in exoplanetary atmospheres, where data may be sparse and noisy, parameterised models are often preferred for their computational speed and flexibility \citep{lineUniformAtmosphericRetrieval2015}. Such models typically specify temperature at discrete pressure levels ($P_i$) with corresponding temperatures $T_i$, and interpolate between these levels using spline functions \citep{lineUniformAtmosphericRetrieval2015}.

To balance physical accuracy with computational efficiency, recent studies have introduced hybrid approaches that combine the physical basis of self-consistent models with the flexibility of free-form models. For instance, \citealt{zhangELementalAbundancesPlanets2023a} proposed a parameterisation that incorporates constraints on temperature gradients from self-consistent models.

In our study, we explored two variations of the temperature parameterisation proposed by \citealt{zhangELementalAbundancesPlanets2023a}. The first model, termed the static gradients (SG) model, utilises a set of temperature gradients ($\nabla_i$) measured at equally spaced pressure levels on logarithmic space ($P_i$). The second model, referred to as the dynamic gradients (DG) model, extends the SG model by allowing the positions of the measured gradients to vary. In the DG model we added three free parameters that determine the pressure level of the RCE boundary ($P_{\text{RCE}}$), the spacing between the upper ($\Delta \log P_{\text{top}}$) and lower pressure levels ($\Delta \log P_{\text{bot}}$). The resulting pressure levels, from the bottom to the top of the atmosphere, are defined as:
\begin{equation}
\begin{aligned}
    P_0 &= 10^2 \text{ bar}, \\
    P_1 &= P_{\text{RCE}} - 2 \Delta \log P_{\text{bot}}, \\
    P_2 &= P_{\text{RCE}} - \Delta \log P_{\text{bot}}, \\
    P_3 &= P_{\text{RCE}}, \\
    P_4 &= P_{\text{RCE}} + \Delta \log P_{\text{top}}, \\
    P_5 &= P_{\text{RCE}} + 2 \Delta \log P_{\text{top}}, \\
    P_6 &= 10^{-5} \text{ bar}.
\end{aligned}
\label{eq:pressure_levels}
\end{equation}

An additional constraint is imposed for the DG model to ensure that the largest temperature gradient occurs at the RCE boundary. The temperature gradients at every layer $\nabla_j$, with $j=0,...,50$, are obtained via linear interpolation of the retrieved parameters $\nabla_i$, with $i=0,...,6$, and the temperature at every layer is calculated according to
\begin{align}
  T_j = T_{j-1} \left(\frac{P_j}{P_{j-1}}\right)^{\nabla_j},
\end{align}
where $T_{j=0}$ is the temperature at the bottom of the atmosphere and is retrieved as a free parameter. This approach enables us to explore a wide range of temperature profiles while maintaining physical consistency and computational efficiency.

\subsubsection{Broadening}\label{subsec:broadening}

Accurately modelling line broadening mechanisms is crucial for HRS. In addition to intrinsic line broadening, the rotation of the object, the turbulence in the atmosphere, and the resolving power of the instrument contribute to the observed line profile.

Rotational broadening is influenced by $v\sin{i}$, where $v$ is the rotational velocity and $i$ is the inclination angle of the rotation axis relative to the line of sight. To account for this, we applied a convolution of the model spectrum with a rotational broadening kernel \citep{grayObservationAnalysisStellar2022}. We used the \texttt{fastRotBroad} function from \texttt{PyAstronomy}\footnote{\url{https://github.com/sczesla/PyAstronomy}} to perform this convolution for each spectral order. This method achieves accuracy within 0.5\% of the more computationally intensive approach of direct disc integration \citep{carvalhoSimpleCodeRotational2023}.

Instrumental broadening accounts for changes in the line profile caused by the Earth's atmosphere and the optical system of the instrument, leading up to the final detector measurement. We modelled this effect using a Gaussian profile, with its standard deviation determined by the full width at half maximum of the telluric lines observed in the spectrum of the standard star (see Sect. \ref{sec:observations}). For spectra of objects with \vsini values greater than the instrumental full width at half maximum, the exact shape of the instrumental line profile has less of an impact, making the use of a Gaussian kernel sufficient for modelling instrumental broadening in such cases.

The total broadening of the model spectrum is obtained by convolution with the rotational broadening kernel and the instrumental line shape. The high resolving power of \CRIRES{} enables precise measurements of the rotational velocity for slow rotators like GQ Lup B. Previous studies \citep{schwarzSlowSpinYoung2016} reported a $v\sin{i} = 5.3^{+1.0}_{-0.9}$ km/s for GQ Lup B using the older CRIRES instrument. Our updated retrievals provide a more precise \vsini measurement (see Eq. \ref{eq:vsini}).

\begin{figure}
  \centering
  \includegraphics[width=\columnwidth]{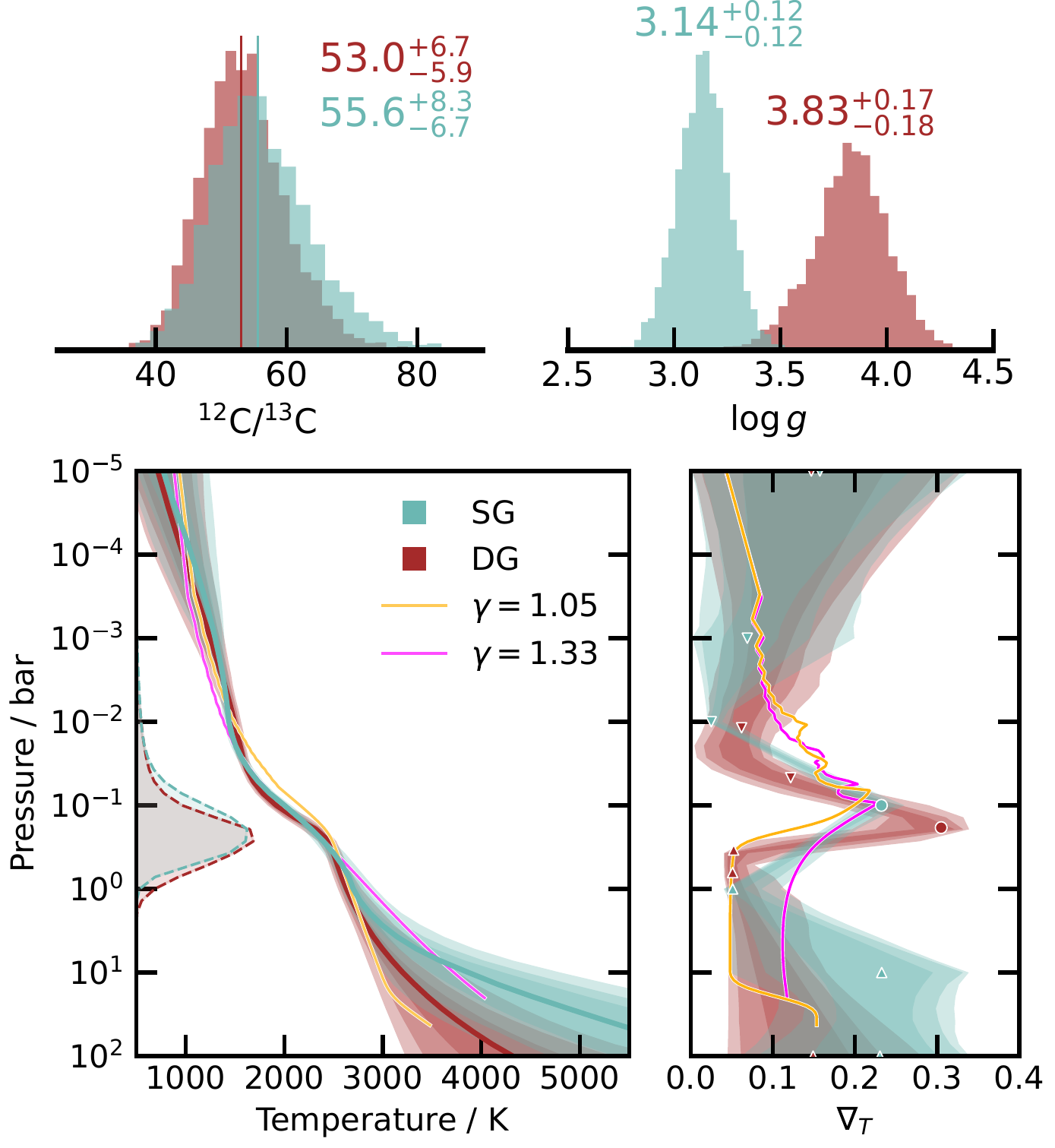}
  \caption{Temperature profile of GQ Lup B retrieved using the SG and DG models. Left: Temperature profile. Right:Temperature gradient. Profiles from the \texttt{ATMO} grid \citep{tremblinFINGERINGCONVECTIONCLOUDLESS2015} with \Teff=2300 K and \logg=4.0 are shown for two distinct models with adiabatic \citep{phillipsNewSetAtmosphere2020} and diabatic convection \citep{petrusMediumresolutionSpectrumExoplanet2021}.}
  \label{fig:fig3_PT}
\end{figure}

\subsection{Telluric transmission}\label{subsec:telluric_transmission}

The observed spectrum contains telluric absorption features originating from the Earth's atmosphere. The contribution on the telluric lines from the stellar signal is already included in the starlight model, as we utilised the observed on-axis spectrum as the basis for the starlight model. However, we must account for the telluric absorption features in the companion spectrum to accurately reproduce the observed spectrum.

To obtain a high-fidelity telluric model, we employed \texttt{Molecfit} to generate synthetic telluric spectra. We utilised the spectrum of a standard star observed immediately before the observations of GQ Lup B to fit a telluric model from \texttt{Molecfit}. The best-fit telluric model corresponds to the atmospheric conditions, including airmass and precipitable water vapour content, at the time of the standard star observations.

To account for changes in the telluric lines between the science and standard star observations, we scaled the telluric model to match the average precipitable water vapour content during the observations of GQ Lup B, which remained approximately constant (refer to \Cref{app:obs_conditions} for details). Additionally, we modelled the scaled best-fit telluric model with one free parameter to fit the effective airmass of the telluric lines in the observed spectrum of GQ Lup B. Saturated telluric lines were masked out during the fitting process, namely we ignored telluric lines with transmission values below 1\% and a region of 20 pixels around them. This procedure aims to accurately incorporate telluric features into our forward model while simultaneously propagating uncertainties in the retrieved parameters.

\subsection{Model comparison}\label{subsec:bayesian_model_comparison}
To find the model that best-fits the spectrum of GQ Lup B, we performed a Bayesian model comparison using the log-likelihood function defined in Eq. \ref{eq:log_likelihood} and we report the goodness of fit of different models in terms of the chi-square statistic \citep{cochranH2TestGoodness1952}. We compared the SG and DG temperature profiles, as well as a model without \thirteenCO{} to evaluate the necessity of including \thirteenCO{} in the model. The Bayesian evidence is calculated as part of the retrieval process, allowing us to calculate the Bayes factor $B_{12}$ between two competing models $\mathcal{M}_1$ and $\mathcal{M}_2$ \citep{kassBayesFactors1995},

\begin{align}
  B_{12} &= \frac{p(\mathbf{d}|\mathcal{M}_1)}{p(\mathbf{d}|\mathcal{M}_2)} = \frac{p(\mathbf{d}|\mathcal{M}_1)}{p(\mathbf{d}|\mathcal{M}_2)} \frac{p(\mathcal{M}_1)}{p(\mathcal{M}_2)}
        &= \frac{Z_1}{Z_2} \frac{p(\mathcal{M}_1)}{p(\mathcal{M}_2)},
\end{align}

where $p(\mathbf{d}|\mathcal{M}_1)$ and $p(\mathbf{d}|\mathcal{M}_2)$ are the marginal likelihoods of models $\mathcal{M}_1$ and $\mathcal{M}_2$, respectively, and $p(\mathcal{M}_1)$ and $p(\mathcal{M}_2)$ are the prior probabilities of the models. The Bayes factor quantifies the relative probability of the models given the data, with a value of $B_{12} > 1$ indicating that model $\mathcal{M}_1$ is more probable than model $\mathcal{M}_2$. Assuming that the prior probabilities of the models are equal, the Bayes factor is equivalent to the ratio of the marginal likelihoods, or in logarithmic form, the difference in the log-evidence values $\ln B_{12} = \log{Z_1} - \log{Z_2}$. The Bayes factor can be converted to a $\sigma$ value using the methodology outlined in \citealt{bennekeHOWDISTINGUISHCLOUDY2013}, providing a quantitative measure of the significance of the model comparison.

\section{Modelling the spectrum of GQ Lup A}\label{sec:modelling_gqlup_a}

The spectrum of the host star, GQ Lup A, is crucial for understanding the properties of the entire system. Here, we focus on modelling the spectrum of GQ Lup A to measure its carbon isotope ratio. Instead of modelling the spectrum with a parameterised model as in the case of GQ Lup B, we use a grid of precomputed PHOENIX models extended to different carbon isotope ratios to fit the observed spectrum of GQ Lup A. This approach allows us to determine the carbon isotope ratio of GQ Lup A while accounting for the veiling effect.

GQ Lup A's \textit{K}-band spectrum is predominantly characterised by CO features and atomic lines. Previous studies have identified it as a young, magnetically active star \citep{donatiMagnetometryClassicalTauri2012}. The presence of strong Zeeman broadening in atomic lines indicates significant magnetic activity, providing opportunities to measure the star's magnetic field strength. For our analysis, we focus on the last two spectral orders, ranging from 2.20 to 2.48 \(\mu m\), where CO lines dominate the spectrum and the influence of Zeeman-broadened atomic lines is small. Additionally, previous high-resolution spectroscopic studies have revealed substantial veiling of photospheric lines in GQ Lup A \citep{sousaNewInsightsNearinfrared2023}, a common feature among young stars \citep{mcclureCHARACTERIZINGSTELLARPHOTOSPHERES2013,reiLinedependentVeilingVery2018}.

\subsection{Isotopic PHOENIX models}\label{subsec:isotopic_phoenix_models} 
Models of stellar atmospheres are a valuable tool for interpreting the observed spectra of stars. The PHOENIX model grid \citep{husserNewExtensiveLibrary2013} provides a wide range of stellar models with varying parameters, including effective temperature, surface gravity, and metallicity. However, the standard PHOENIX grid does not include models with different carbon isotope ratios. To address this limitation, we generated a custom grid of PHOENIX models with varying \Cratio{} from 16 to 301 in steps of 15. For our purposes, we used models with fundamental parameters similar to GQ Lup A (see \Cref{tab:prop_gqlup}). 
\begin{figure}
    \centering
    \includegraphics[width=\columnwidth]{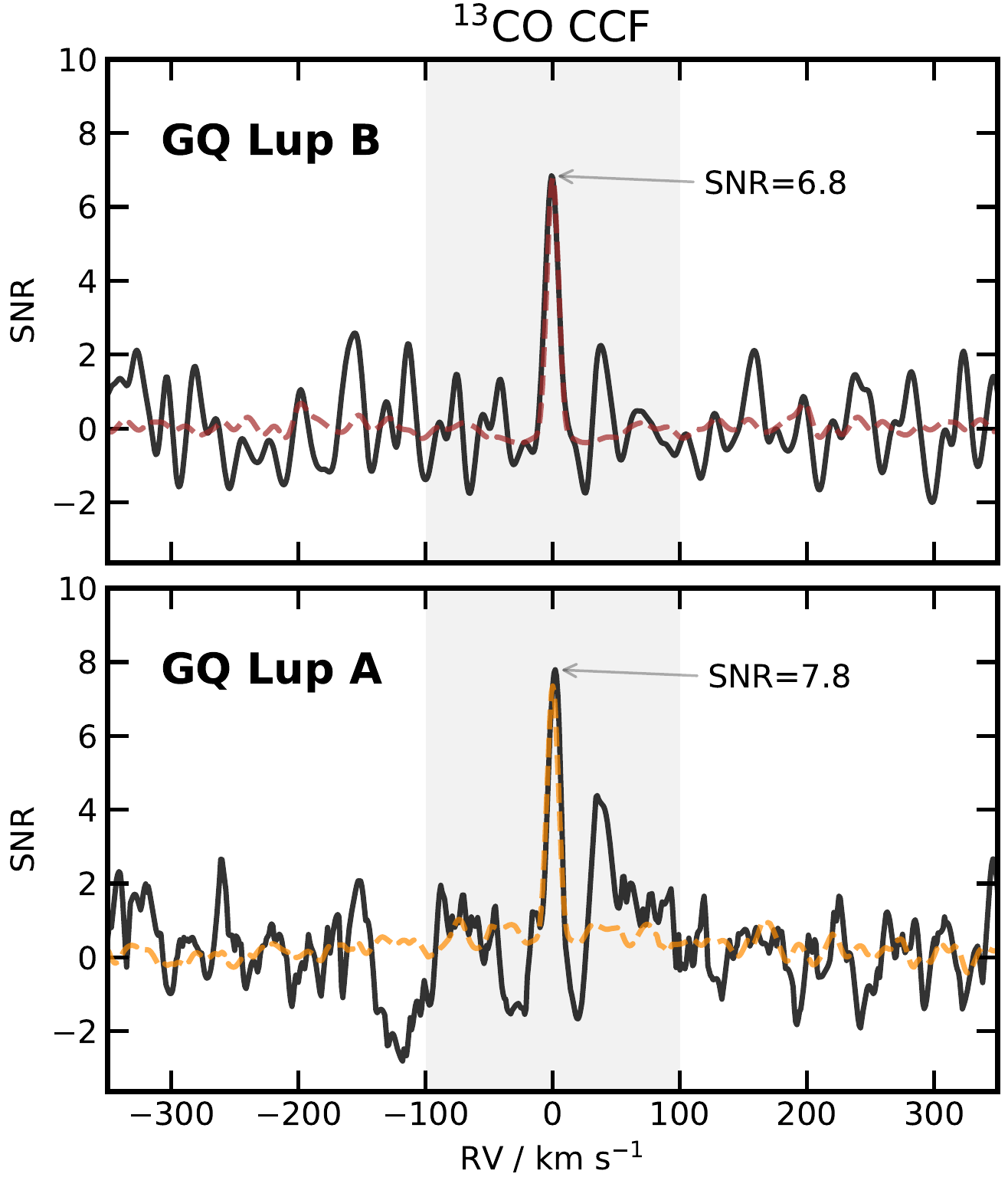}
    \caption{Cross-correlation functions (CCFs) of the data and the best-fit models of \thirteenCO{} for GQ Lup B (top) and GQ Lup A (bottom). The CCFs are shown in black, with the auto-correlation functions of the templates shown in brown and orange. The CCFs are calculated between the data residuals and the model residuals following the methodology from \citealt{zhang13COrichAtmosphereYoung2021} (also in \citealt{deregtESOSupJupSurvey2024}).}
    \label{fig:ccf_AB}
  \end{figure}

\begin{figure*}
  \centering
  \begin{subfigure}[t]{0.72\textwidth}
    \centering
    \includegraphics[width=\textwidth]{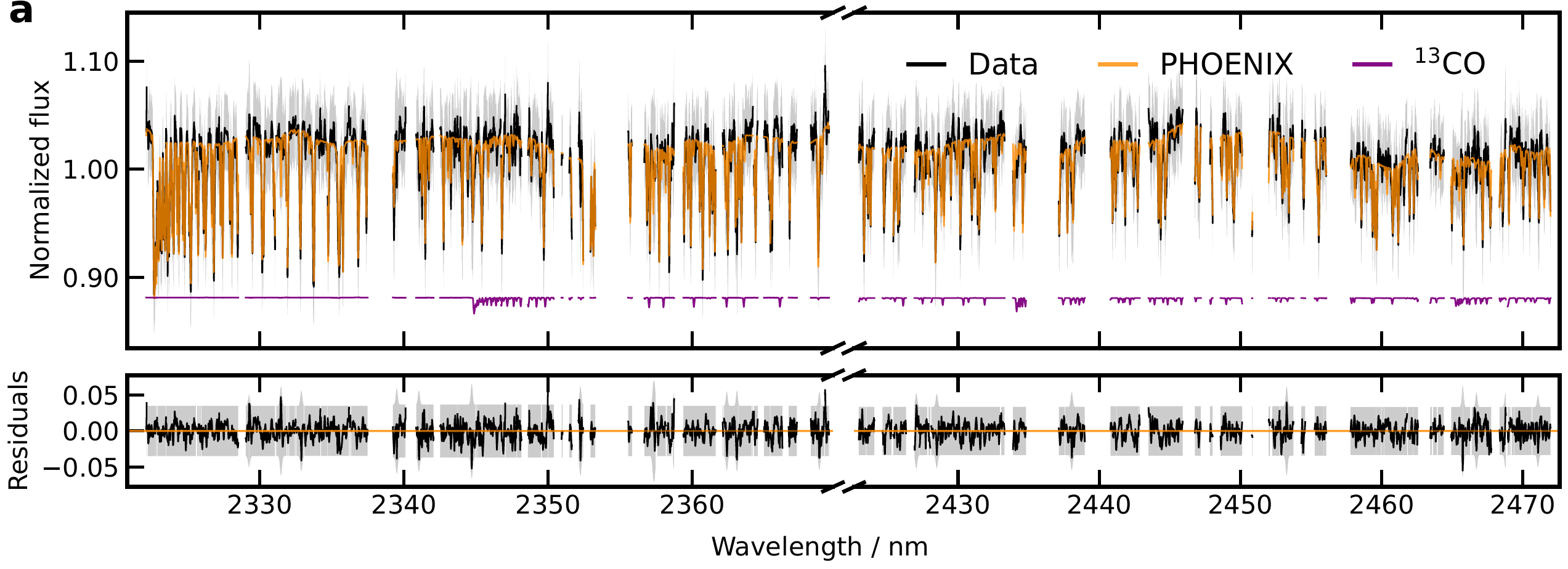}
    \label{fig:fig4_bestfit_spec}
  \end{subfigure}
  \hfill
  \begin{subfigure}[t]{0.27\textwidth}
    \centering
    \includegraphics[width=\textwidth]{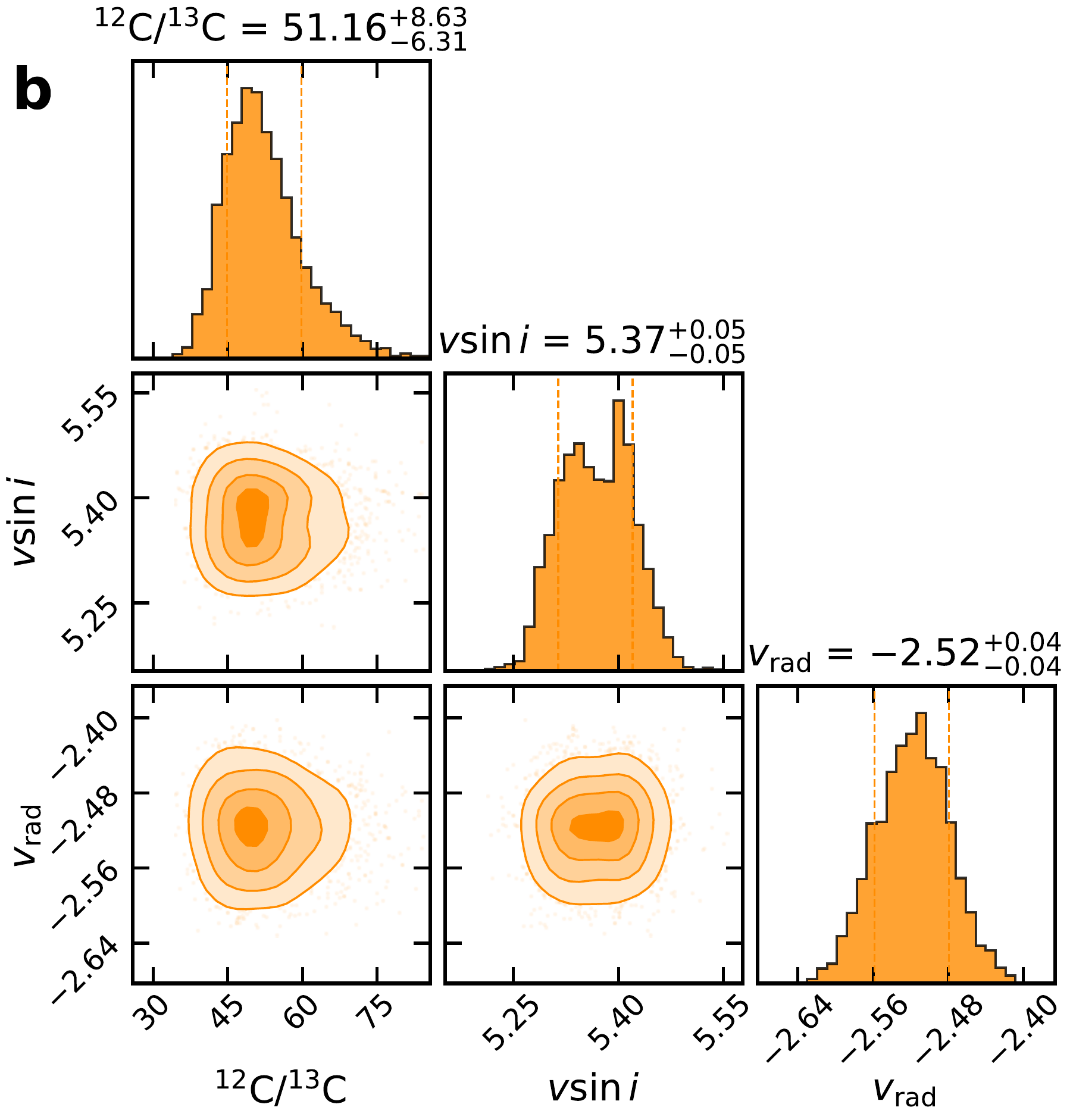}
    \label{fig:fig4_corner_plot}
  \end{subfigure}
  \caption{Best-fit model of GQ Lup A \textbf{a}. Spectrum with the veiling continuum subtracted and divided by the telluric model. The observed spectrum is shown in black, the PHOENIX model with optimal linear amplitudes in orange and the \thirteenCO{} lines from the best-fit model in purple. The residuals of the are shown in the bottom panel, the shaded grey region indicates the root-squared diagonal of the covariance matrix with the GP kernel. \textbf{b}. Posterior distributions of the carbon isotope ratio, projected rotational velocity and radial velocity of GQ Lup A.}
  \label{fig:fig4_bestfit_star}
\end{figure*}

\subsection{Linear veiling model}\label{subsec:linear_model_gqlup_a}

Veiling reduces the depth of absorption lines, making them appear shallower than those of a main sequence star of the same spectral type \citep{hartiganHowUnveilTauri1989}, with the veiling factor (\(r_k\)) typically increasing with wavelength \citep{fischerCHARACTERIZINGIYJEXCESS2011}. To model the spectrum of GQ Lup A, we used a linear combination of a PHOENIX model for the star ($M_A$) and a veiling model parameterised as a slowly varying function of wavelength ($M_{\text{veiling}}$). The veiling component was fitted to the data using a spline model with ten knots, following a similar approach used for the starlight model. The linear model is expressed as:

\begin{align}
  d &= \phi M + n = \phi_{A} M_{A} + \phi_{\text{veiling}} M_{\text{veiling}} + n.
\end{align}

During the likelihood calculation, the amplitude of the linear parameters is fitted, providing a robust estimation of each source's contribution. The veiling factor \( r_k \), defined as the ratio of the veiling amplitude to the amplitude of the model spectrum of GQ Lup A, is calculated as:

\begin{align}
  r_k = \frac{\phi_{\text{veiling}}}{\phi_{A}}.
\end{align}

To determine the best-fit model of the spectrum of GQ Lup A, we fitted the observed spectrum with the model spectrum of GQ Lup A and the veiling model. We set up a retrieval framework with six free parameters, covering the stellar properties (radial velocity, spin, and carbon isotope ratio), the effective airmass of the observations, and the parameters of the GP kernel. The likelihood function is defined similarly to Eq. \ref{eq:log_likelihood}, with the amplitude of the PHOENIX model and the veiling model as the linear components. 

\subsection{Retrieval of GQ Lup A}\label{subsec:retrieval_gqlup_a}
We retrieved the best-fit parameters of GQ Lup A using the linear model described in \Cref{subsec:linear_model_gqlup_a}. To avoid degeneracies between veiling and the stellar parameters, we fixed the effective temperature to \Teff=4300 K as reported in \Cref{tab:prop_gqlup}. In addition, the surface gravity and metallicity were also fixed to \logg=4.0 and [Fe/H]=0.0, respectively. We modelled the noise of the observed spectrum using a GP kernel following the approach outlined in \Cref{subsec:correlated_noise}. The model spectra were broadened according to the instrumental resolution of \CRIRES{} and the rotational velocity \vsini of GQ Lup A, which we fitted for as part of the retrieval process. Additionally, we forward-modelled the telluric transmission spectrum by allowing the airmass of the observations to vary as a free parameter.

\section{Results}\label{sec:results}

\subsection{The atmosphere of GQ Lup B}
The atmospheric retrieval process provides us with a best-fit model characterising the atmosphere of GQ Lup B, along with its rotational and radial velocities. This model effectively reproduces observed spectral features across the entire wavelength range up to the reported uncertainty level. However, the derived surface gravity of GQ Lup B faces challenges due to degeneracies with the temperature profile, resulting in less constrained estimates. We present the main results from the retrieval with the DG model, which is favoured over the SG with a Bayesian evidence of $B_{\text{DG/SG}} = 8.53$, equivalent to a 4.5$\sigma$ preference \citep{bennekeHOWDISTINGUISHCLOUDY2013}.

\subsubsection{Chemical composition}\label{subsec:results_chemical_composition}
\Cref{tab:free_params} summarises the abundances of the primary opacity sources in GQ Lup B's atmosphere. From the retrieved abundances of \water, \twelveCO{} and \thirteenCO{} we derived the C/O ratio, 
\begin{align}
    \rm C/O = 0.50^{+0.01}_{-0.01},
\end{align}
which is comparable to the solar value of C/O = 0.59 $\pm$ 0.08 \citep{asplundChemicalMakeupSun2021}. The presence of \thirteenCO{} is confirmed at $5.8\sigma$ and clearly detected in the cross-correlation with the best-fit model (S/N=6.8; see the top panel of \Cref{fig:ccf_AB}). The derived  \Cratio{} from the retrieved abundances of \twelveCO{} and \thirteenCO{} is
\begin{align}
    \mathrm{^{12}C/^{13}C} = 53^{+7}_{-6},
\end{align}
which is broadly consistent with the lower value of the local ISM of \Cratio{} = 69 $\pm$ 15 \citep{milam1213Isotope2005} and lower than the Sun's value of \Cratio{} = $93.5 \pm 3.1$ \citep{lyonsLightCarbonIsotope2018}. 

We report the metallicity of GQ Lup B as the carbon-to-hydrogen ratio normalised to the solar value as defined in Eq. \ref{eq:metallicity},

\begin{align}
    \mathrm{[C/H]} = 0.50^{+0.16}_{-0.17},
\end{align}

we note that [C/H] is strongly degenerate with \logg{}, exhibiting a Pearson correlation coefficient of $r=0.97$\footnote{\href{https://docs.scipy.org/doc/scipy/reference/generated/scipy.stats.pearsonr.html}{scipy.stats.pearsonr}}, which indicates that the retrieved metallicity is highly dependent on the surface gravity of the object. In the best-fit model with the SG temperature profile, the retrieved metallicity is [C/H] = $-0.12^{+0.10}_{-0.10}$, with $r=0.95$, corresponding to the lower \logg{} retrieved in this model.

We constrained the abundances of HF, Ca, Na and Ti in the atmosphere of GQ Lup B, with values comparable to those reported for three young BDs with similar spectral types, ages, and temperatures to GQ Lup B \citep{gonzalezpicosESOSupJupSurvey2024}.
We report the non-detection of the following species: \eighteenOwater, C${}^{18}$O, Mg, K and Fe. Considering the temperature and wavelength range of the observations, no additional molecules are expected to be relevant sources of opacity in the atmosphere \citep{iyerSPHINXMdwarfSpectral2023}.

\subsubsection{Temperature profile}\label{subsec:results_temperature_profile}

The retrieved temperature profile and its gradient are shown in the bottom panels of \Cref{fig:fig3_PT}. Both pressure-temperature (PT) parameterisations are consistent in the upper atmosphere and near the radiative-convective boundary. However, slight differences arise in the deeper atmospheric layers.

The more complex DG model is favoured over the simpler SG model, as supported by both frequentist and Bayesian analyses. From the frequentist perspective, the chi-square difference is \(\Delta \chi^2 = 11.24\), which exceeds the critical value of \(9.49\) at the 5\% significance level (\(\alpha = 0.05\)) and corresponds to a $p$-value of \(0.024\). Bayesian evidence further supports this preference, with a Bayes factor of \(B_{\text{DG/SG}} = 8.53\), equivalent to a 4.5$\sigma$ preference \citep{bennekeHOWDISTINGUISHCLOUDY2013}. We note that the retrieved parameters of the covariance matrix are identical ($<1\sigma$) for the two models, and hence the Bayesian evidence comparison should not be affected by the covariance matrix. These results underscore the DG model’s improved fit and greater flexibility in capturing the temperature profile.

The retrieval constrains the position and temperature gradient of the RCE boundary in GQ Lup B's atmosphere, enabling the derivation of $T_{\text{RCE}}$,

\begin{align}
    \begin{aligned}
        P_\mathrm{RCE} &= 0.19^{+0.02}_{-0.02}\ \mathrm{bar},\\
        \nabla_{\text{RCE}} &= 0.32^{+0.01}_{-0.02},\\
        T_\mathrm{RCE} &= 2319.91^{+36.28}_{-43.80}\ \mathrm{K},
    \end{aligned}
\label{eq:RCB_values}
\end{align}

this highlights the pressure and temperature range of the atmosphere that our observations probe. The temperature gradient at the RCE boundary aligns with the expected value for the adiabatic cooling of a hydrogen and helium gas mixture ($\nabla_{\text{ad}} \approx 0.28-0.38$; \cite{coxPrinciplesStellarStructure1968,hansenStellarInteriorsPhysical2004}).

\subsubsection{Rotational and radial velocity}\label{subsec
}
The best-fit projected rotational velocity of GQ Lup B is 
\begin{align}
    v\sin{i} &= 5.56^{+0.02}_{-0.02}\ \mathrm{km/s}, \label{eq:vsini}
\end{align}
which is consistent with the previous measurement of $v\sin{i} = 5.3^{+1.0}_{-0.9}$ km/s using the old CRIRES \citep{schwarzSlowSpinYoung2016}, confirming a slow spin characteristic of its young age. The retrieved radial velocity of GQ Lup B, corrected for barycentric motion is
\begin{align}
    v^{B}_\mathrm{rad} &= -0.49^{+0.02}_{-0.02}\ \mathrm{km/s} - v_\mathrm{sys} = 2.03^{+0.04}_{-0.04}\ \mathrm{km/s}. \label{eq:RV_corr}
\end{align}

where we have used the radial velocity of the host star as the systemic velocity, $v_\mathrm{sys} = -2.52^{+0.04}_{-0.04}$ km/s (see Eq. \ref{eq:RV_A}). Our radial velocity measurement is in good agreement with the earlier measurement of $2.0 \pm 0.4$ km/s  by \citealt{schwarzSlowSpinYoung2016}. We note that our reported uncertainty does not account for potential variability due to magnetospheric accretion as reported in the literature, which could introduce an additional uncertainty source of approximately  $0.4$ km/s \citep{donatiMagnetometryClassicalTauri2012}.

\subsubsection{Surface gravity}\label{subsec:surface_gravity}
The surface gravity of the best-fit model of GQ Lup B is
\begin{align}
  \logg = 3.83^{+0.17}_{-0.18}.
\end{align}
This value should be interpreted with caution as it is subject to degeneracies with the temperature profile and the absolute abundances or metallicity. We performed a test retrieval applying a high-pass filter to the data and the model to remove the continuum, which led to similar results to the nominal retrieval.

\subsection{The spectrum of GQ Lup A}\label{subsec:results_gqlup_a}
The veiling model effectively captures the continuum excess and the decreased line depths present in the observed spectrum, providing a robust estimation of the veiling factor at each wavelength and enabling the retrieval of the isotope ratio with PHOENIX models. The best-fit model of GQ Lup A is shown in \Cref{fig:fig4_bestfit_star}, with the residuals indicating a good fit to the data to within $<5\%$. The contribution from the veiling continuum and the PHOENIX model are shown in \Cref{app:veiling_gqlup_a}.

\subsubsection{Veiling factor}\label{subsec:veiling_factor}
\begin{figure}[h!]
  \centering
  \includegraphics[width=\columnwidth]{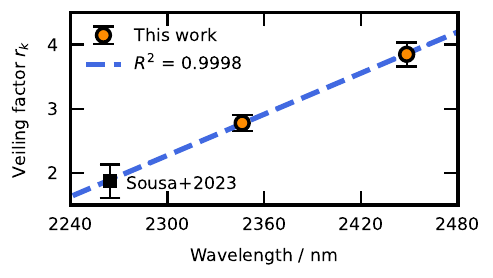}
  \caption{Veiling factor of GQ Lup A as a function of wavelength. The veiling factors for each spectral order derived from the fitted linear amplitudes are shown in orange, with error bars representing the standard deviation between the three detectors of each order. The measurement from \citealt{sousaNewInsightsNearinfrared2023} is included to fit the three points with a linear function of slope 0.011 nm$^{-1}$.}
  \label{fig:veiling_factor}
\end{figure}

The excess continuum observed in the spectrum of GQ Lup A dominates over the PHOENIX continuum. The strength of veiling increases with wavelength, with a veiling factor of $2.78 \pm 0.13$ and $3.85 \pm 0.19$ at 2346 nm and 2448 nm, respectively. In \Cref{fig:veiling_factor} we show that the veiling factor increases linearly with wavelength over the observed range, consistent with a previous measurement at 2260 nm \citep{sousaNewInsightsNearinfrared2023}. 

\subsubsection{Rotational and radial velocity}
The retrieved projected rotational velocity of GQ Lup A 
\begin{align}\label{eq:vsini_A}
  v\sin{i} = 5.37 \pm 0.05 \text{ km/s},
\end{align}
 indicates a slow rotator, in line with expected values for young stars \citep{bouvierAngularMomentumEvolution2014}. Previous measurements of the rotational velocity of GQ Lup A have reported values ranging from 5 to 7 km/s \citep{guentherLowmassCompanionGQ2005,donatiMagnetometryClassicalTauri2012,schwarzSlowSpinYoung2016}. The presence of stellar spots and magnetic activity in GQ Lup A can lead to variations in the observed \vsini, which may explain the range of values reported in the literature. \citealt{donatiMagnetometryClassicalTauri2012} conducted a detailed study of the magnetic field of GQ Lup A on spectropolarimetric data and reported a rotational velocity of $v\sin{i} = 5.0 \pm 1.0$ km/s. Our measurement is consistent with this value, thus providing a refined measurement of the projected rotational velocity of GQ Lup A. We report a radial velocity of 
\begin{align}\label{eq:RV_A}
  v_{\text{rad}} = -2.52 \pm 0.04 \text{ km/s}
\end{align}

 which is broadly consistent with previous measurements of the systemic velocity \citep{donatiMagnetometryClassicalTauri2012,schwarzSlowSpinYoung2016}.

\subsubsection{Carbon isotope ratio}
The best-fit carbon isotope ratio for GQ Lup A is
\begin{align}
\mathrm{^{12}C/^{13}C} = 51^{+10}_{-8},
\end{align}
with uncertainties that include a systematic error of $\Delta$\Cratio=5, arising from the level of telluric masking (0.10-0.50) and the selection of \Teff{} for the PHOENIX model within the 4200-4400 K range. The significance of this measurement is underscored by the Bayes factor, comparing models with and without \thirteenCO{} in the PHOENIX grid, resulting in $\ln B{\text{with/without}} = 15.1$, equivalent to a 5.8$\sigma$ preference for the inclusion of \thirteenCO{}. Additionally, the cross-correlation function of the residuals between the data and the GQ Lup A model further corroborates a strong detection of \thirteenCO{} (S/N=7.8; see the bottom panel of \Cref{fig:ccf_AB}).

\section{Discussion}\label{sec:discussion}

\subsection{Comparison with previous studies}
The analysis of the KPIC spectrum of GQ Lup B, as presented in \citealt{xuanAreThesePlanets2024} (hereafter X24), reports a \Cratio{} value for GQ Lup B that is inconsistent with our findings at a significance level of $\geq3\sigma$:
\begin{align*}
  &\rm \Cratio_{\text{X24}} = 153^{+43}_{-31},\\
  &\rm \Cratio_{\text{this work}} = 53^{+7}_{-6},
\end{align*}

both studies confirm the presence of \thirteenCO{} in the atmosphere of GQ Lup B; however, the derived \Cratio{} values differ significantly. Differences in wavelength coverage and spectral resolution between KPIC and \CRIRES{} likely contribute to this discrepancy. X24 employs the three reddest spectral orders of KPIC, spanning wavelength ranges of (2290–2340 nm), (2360–2410 nm), and (2440–2490 nm) at a resolution of $\mathcal{R} \sim 35,000$. In contrast, our analysis utilises the full wavelength range of \CRIRES{} K2166, covering seven spectral orders between 1920 nm and 2472 nm (see detailed ranges in \Cref{fig:bestfit_all_orders}) at $\mathcal{R} \sim 120,000$. Furthermore, our observations achieve a higher signal-to-noise ratio (S/N), with an estimated S/N $\sim 20$ for the companion spectrum compared to S/N $\sim 12$ for KPIC. Our retrieval independently fits the abundances of individual species without enforcing equilibrium chemistry, whereas X24 derives species abundances from free parameters of C/O and metallicity under the assumption of chemical equilibrium.

X24 reports a super-adiabatic temperature gradient around the photosphere of GQ Lup B, exceeding predictions from RCE (see \Cref{fig:corner_fixed_PT}). In contrast, our retrieved temperature profile shows a high photospheric gradient of approximately 0.32, though significantly lower than 0.50 as retrieved by X24, and we do not find evidence of a super-adiabatic region. These discrepancies in temperature profiles may contribute to differences in the derived abundances and \Cratio{} values, as discussed in \Cref{subsec:fixed_PT}.

The C/O ratio retrieved by X24 ($0.70 \pm 0.01$) is higher than the solar value and our result ($0.50 \pm 0.01$). A higher abundance of \twelveCO{} could inflate both the C/O and \Cratio{} values while maintaining a constant \thirteenCO{} abundance. Our independent fits to the main opacity sources allow us to avoid assumptions of chemical equilibrium, providing an alternative perspective. Further studies are required to reconcile these differences, including a detailed comparison of individual abundances and the interaction between the pressure-temperature profile and surface gravity.

\subsection{Atmospheric retrieval with fixed thermal profile from Xuan et al. (2024)}\label{subsec:fixed_PT}
The absorption line depths of atmospheric species are sensitive to the temperature profile at different altitudes. Spectral lines of different species form at distinct atmospheric layers depending on the temperature structure and opacity. As a result, species probing different line depths sample varying pressure regions. The \Cratio{} is particularly sensitive to the temperature profile because \thirteenCO{} lines originate deeper in the atmosphere than \twelveCO{} lines. These regions can be identified using the contribution function \citep{mollierePetitRADTRANSPythonRadiative2019}, which calculates the relative line contrast of a species across pressures, indicating relevant altitudes where absorption lines originate (see \Cref{fig:contr_em_CO}).

To evaluate whether the temperature profiles retrieved in this work and X24 explain the \Cratio{} discrepancy, we performed a retrieval using the fixed temperature profile from X24 (FPTX24). The posterior distributions for our nominal retrieval (DG model) and FPTX24 are shown in \Cref{fig:corner_fixed_PT}. A preference for the DG model is observed, with an 8.6$\sigma$ improvement in fit quality over FPTX24. The FPTX24 retrieval yields systematically lower abundances and a reduced surface gravity:
\begin{align*}
    \logg_{\text{FPTX24}} = 2.28^{+0.14}_{-0.12} < 3.83^{+0.17}_{-0.18} = \logg_{\text{DG}}.
\end{align*}

The lower surface gravity in FPTX24 stems from differences in the temperature profile. The \Cratio{} retrieved with FPTX24 ($68^{+12}_{-10}$) is higher than our nominal value ($53^{+7}_{-6}$) but consistent within uncertainties. Similarly, the FPTX24 C/O ratio ($0.57 \pm 0.02$) exceeds the DG model's value ($0.50 \pm 0.01$). These results suggest that distinct temperature profiles impact species abundances and spectral interpretations. However, the extent of the discrepancy does not fully account for the difference in \Cratio{} values reported by X24 and this work.

\subsection{The carbon isotope ratio of the GQ Lup system}\label{subsec:carbon_isotope_ratio}
The \Cratio{} has been proposed as a potential tracer of formation pathways of BDs and SJs. In the presence of ice accretion, giant planets might be enriched in \Cratio{} compared to its host-star \citep{zhang13COrichAtmosphereYoung2021}. The measured \Cratio{} ratios in the GQ Lup system suggest that GQ Lup B is not significantly enriched in ${}^{13}$C compared to GQ Lup A. This result is consistent with the formation of GQ Lup B through gravitational collapse or disc instability \citep{stamatellosPropertiesBrownDwarfs2009,stolkerCharacterizingProtolunarDisk2021}. The \Cratio{} of GQ Lup A/B is slightly lower than the local ISM $\sim 69$ \citep{wilsonIsotopesInterstellarMedium1999} but broadly consistent within the 1$\sigma$ scatter \Cratioismilam{} \citep{milam1213Isotope2005}. The measured \Cratio{} of GQ Lup B lies between the reported \thirteenCO{} atmosphere of YSES 1 b ($30^{+10}_{-7}$; \citealt{zhang13COrichAtmosphereYoung2021}) and VHS 1256 b ($62\pm 2$; \citealt{gandhiJWSTMeasurements132023}). Additional measurements of the \Cratio{} of young SJs and BDs are required to understand the diversity of formation pathways and the role of \Cratio{} in the formation of substellar objects.

To assess the role of \Cratio{} as a formation tracer it is critical to measure the \Cratio{} of the host star and the companion. A high \thirteenCO{} abundance can be interpreted as a signature of ice accretion during the formation of the object \citep{zhang13COrichAtmosphereYoung2021}; however, it should be compared to the \Cratio{} of the host star to understand the formation history of the system and whether there is any actual enrichment due to accretion processes. Recent work by \citealt{xuanValidationElementalIsotopic2024} presented a homogeneous isotopic composition for the binary system HIP 555077, where the \Cratio{} of the host star and the companion are consistent with each other. We report a similar consistency in the \Cratio{} of GQ Lup A and GQ Lup B, suggesting that the system formed from the same parent cloud that was somewhat enriched in \thirteenCO{} compared to the present-day local ISM. This result is cohesive with the view that the \Cratio{} of young objects should have a higher abundance of \thirteenCO{} compared to older objects due to the enrichment of \thirteenCO{} in the ISM over time (e.g. \citealt{prantzosEvolutionCarbonOxygen1996,milam1213Isotope2005}). In the first results of the SupJup survey, \citealt{deregtESOSupJupSurvey2024} reported the carbon isotope ratio of an old T dwarf (DENIS 0255, \Cratio=$184^{+40}_{-61}$), which is signifcantly higher than the ISM. On the other hand, the \Cratio{} of three young BDs (age $\leq 10$ Myr) as presented in \citealt{gonzalezpicosESOSupJupSurvey2024} are consistent with the ISM, with values ranging from $\approx 80-120$. 

\subsection{The carbon-to-oxygen ratio}\label{subsec:C_O_ratio}

The carbon-to-oxygen ratio (C/O) has been proposed as a chemical indicator of formation pathways for substellar objects (e.g. \citealt{obergEFFECTSSNOWLINESPLANETARY2011}, \citealt{fortneyCARBONTOOXYGENRATIOMEASUREMENT2012}, \citealt{hochAssessingRatioFormation2023}). Objects formed within protoplanetary discs may exhibit a C/O ratio exceeding the solar value (around 0.59) due to water freeze-out onto dust grains, enriching the disc's gas-phase C/O (\citealt{obergEFFECTSSNOWLINESPLANETARY2011,mordasiniIMPRINTEXOPLANETFORMATION2016,brewerRATIOSSUGGESTHOT2017,cridlandConnectingPlanetFormation2019}). Conversely, a C/O ratio closer to the solar value might suggest formation through mechanisms like gravitational collapse or disc instability, where the C/O reflects the bulk composition of the parent cloud (\citealt{bossFORMATIONGIANTPLANETS2011}).

However, interpreting atmospheric C/O in substellar objects remains challenging due to potential condensation of oxygen-bearing molecules (e.g. \citealt{lineUniformAtmosphericRetrieval2015}, \citealt{calamariAtmosphericRetrievalBrown2022}). This effect is particularly pronounced for cooler objects than GQ Lup B, such as VHS 1256 b \citep{gandhiJWSTMeasurements132023}, where a fraction of the oxygen is expected to be sequestered in clouds, leading to an enhanced C/O ratio in the gas phase.

\begin{figure}
  \centering
  \includegraphics[width=\columnwidth]{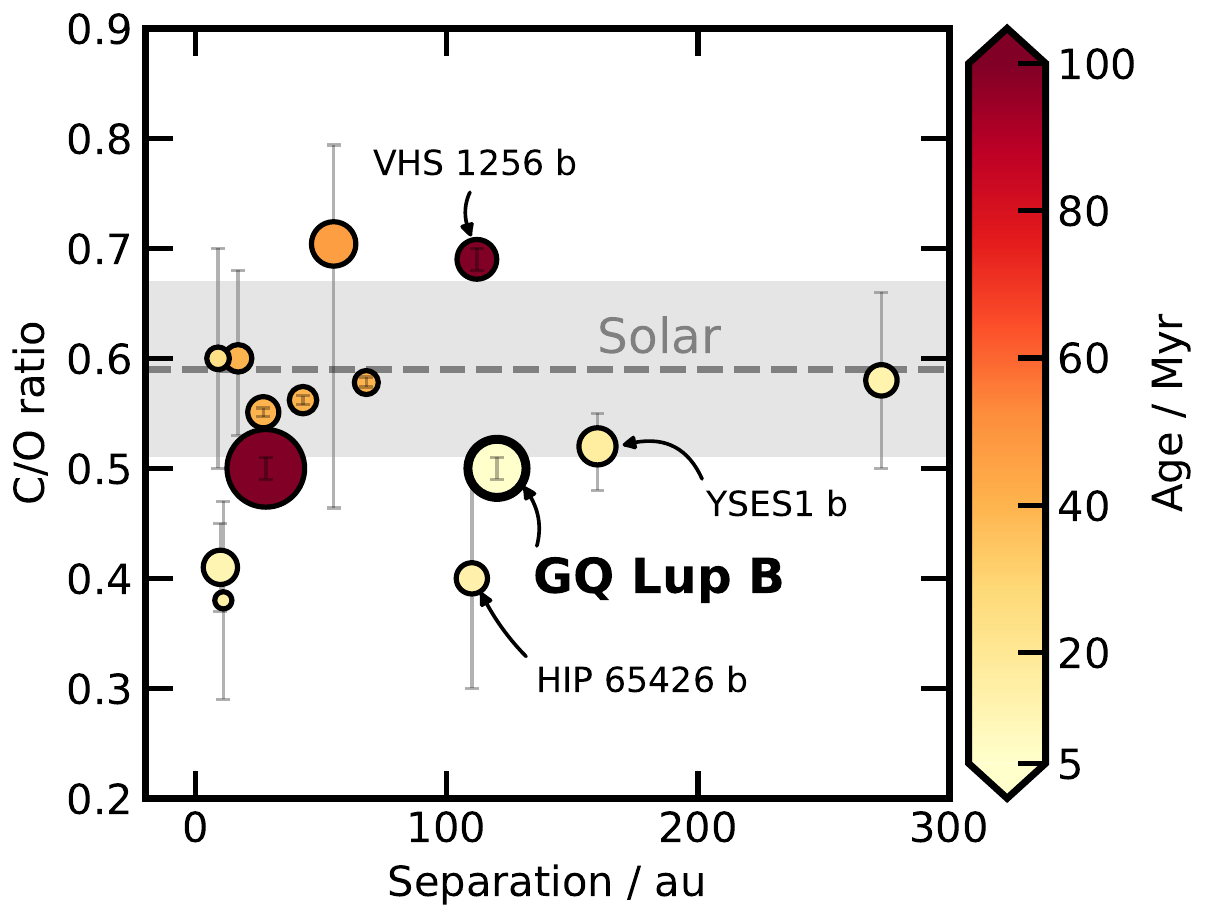}
  \caption{Carbon-to-oxygen ratio of GQ Lup B compared to other young substellar companions ($<100$ Myr). The solar value and its uncertainty are shown as a grey band. The size of the markers is proportional to the mass. The data used to create this figure and their uncertainties can be found in 
\citealt{petrusMediumresolutionSpectrumExoplanet2021}, \citealt{zhang13COrichAtmosphereYoung2021}, \citealt{lacourMassPictorisPictoris2021}, \citealt{dupuyMassesAgeArchitecture2023}, \citealt{gandhiJWSTMeasurements132023}, \citealt{hochAssessingRatioFormation2023}, \citealt{brown-sevillaRevisitingAtmosphereExoplanet2023}, \citealt{mesaAFLepLowest2023}, \citealt{palma-bifaniPeeringYoungPlanetary2023}, \citealt{palma-bifaniAtmosphericPropertiesAF2024}, \citealt{landmanPictorisEyesUpgraded2024} and \citealt{costesFreshViewHot2024}.}
  \label{fig:CO_ratio}
\end{figure}

Our analysis reveals a C/O ratio for GQ Lup B of $0.50 \pm 0.01$, consistent with previous measurements from medium-resolution spectroscopy ($0.44^{+0.13}_{-0.11}$; \citealt{demarsEmissionLineVariability2023}). This finding aligns with a formation scenario via gravitational collapse or disc instability for GQ Lup B. Furthermore, our measurement for GQ Lup B falls within the range observed for other young substellar objects (see \Cref{fig:CO_ratio}), such as YSES 1 b ($0.52^{+0.04}_{-0.03}$; \citealt{zhang13COrichAtmosphereYoung2021}) and HD 984 B ($0.50^{+0.01}_{-0.01}$; \citealt{costesFreshViewHot2024}). We note that the reported value from \citealt{xuanAreThesePlanets2024} ($0.70\pm0.01$) is higher than the one reported here, as discussed in the previous section, this may be attributed to a different model interpretation. While C/O alone may not be a definitive formation tracer, it provides valuable insights into the atmospheric chemistry and complements isotope ratio studies in piecing together the formation history of substellar objects.

\subsection{Non-adiabatic convection in GQ Lup B}\label{subsec:non_adiabatic}
The retrieved temperature profile of GQ Lup B shows a slight deviation from the adiabatic temperature gradient in the lower atmosphere, suggesting the presence of non-adiabatic convection \citep{tremblinFINGERINGCONVECTIONCLOUDLESS2015,tremblinThermocompositionalDiabaticConvection2019,petrusMediumresolutionSpectrumExoplanet2021}. This result underscores the importance of considering non-adiabatic convection in the interpretation of spectra of SJs and BDs, as it can significantly affect the temperature structure and the derived abundances. The presence of non-adiabatic temperature profiles has been reported in other objects, such as the cool Y dwarf WISE 0359 \citep{leggettFirstDwarfData2023}.

\subsection{Sensitivity of the surface gravity to the temperature profile}\label{subsec:sensitivity_surface_gravity}

Atmospheric retrievals using different parameterisations of the PT profile reveal that small differences in temperature gradients can significantly influence the retrieved surface gravity. The DG model offers a more flexible representation of the atmospheric temperature structure by allowing the photosphere's position to be fitted, avoiding biases associated with predefined pressure levels. In contrast, the SG model constrains the surface gravity to \(\log{g} = 3.14 \pm 0.12\), whereas the DG model retrieves a higher surface gravity of \(\log{g} = 3.83 \pm 0.18\).
The DG model is preferred due to its more adaptable temperature structure and superior fit to the data, as evidenced by a 4.5$\sigma$ preference and a lower reduced chi-squared value. The surface gravity retrieved with the DG model aligns well with previous literature measurements, suggesting it better captures the true atmospheric temperature structure of GQ Lup B \citep{seifahrtNearinfraredIntegralfieldSpectroscopy2007,lavigneNEARINFRAREDOBSERVATIONSGQ2009}, also consistent with the expected surface gravity for young substellar objects \citep{baraffeEvolutionaryModelsLowmass2002}. 

Using evolutionary models, \citet{stolkerCharacterizingProtolunarDisk2021} and \citet{xuanAreThesePlanets2024} estimated a surface gravity of \(\log{g} \approx 3.8\). While this value agrees with our retrieval, uncertainties in the system's age and limitations in evolutionary models may introduce discrepancies in the predicted surface gravity.

\subsection{Veiling of the GQ Lup A spectrum}
The \textit{K}-band veiling factor of GQ Lup A is consistent with a rapid increase with wavelength including the measurement from \citealt{sousaNewInsightsNearinfrared2023} (see \Cref{fig:veiling_factor}). The veiling contribution is an essential parameter for the accurate modelling of the spectrum of GQ Lup A, as it affects the depth of the absorption lines (see \Cref{fig:gqlupa_bestfit_linear_model}). 

Previous attempts to link the observed veiling to disc emission have found black body temperatures that are too high for dust to survive in the inner disc \citep{fischerCHARACTERIZINGIYJEXCESS2011,alcalaGIARPSHighresolutionObservations2021}. Alternative descriptions of veiling invoke the presence of hot spots on the stellar surface or accretion emission \citep{hartiganHowUnveilTauri1989,mcclureCHARACTERIZINGSTELLARPHOTOSPHERES2013,kidderIGRINSYSOSurvey2021}. Observations at longer wavelengths (e.g JWST/MIRI; \citealt{cugnoMidInfraredSpectrumDisk2024}) might be able provide insights into the origin of veiling in GQ Lup A.

\section{Conclusions}\label{sec:conclusions}

In this study we analysed the carbon isotope ratio (\Cratio{}) in the GQ Lup system and the carbon-to-oxygen ratio (C/O) of its companion, GQ Lup B, to investigate their formation pathways. Our findings suggest that GQ Lup B formed through gravitational collapse or disc instability, as indicated by the homogeneous \Cratio{} of the system and the lack of enhanced carbon elemental abundance. The key conclusions of this study are as follows:
\begin{itemize}
    \item Carbon isotope ratio: The \Cratio{} of GQ Lup B ($\mathrm{^{12}C/^{13}C} = 53^{+7}_{-6}$) is consistent with that of GQ Lup A ($\mathrm{^{12}C/^{13}C} = 51^{+10}_{-8}$), indicating a shared origin from the same parent cloud with no evidence of \thirteenCO{} enrichment. This supports the idea of formation through gravitational collapse or disc instability, rather than significant solid accretion.

    \item Comparison with other systems: The \Cratio{} of GQ Lup B falls within the range observed for other young substellar objects but is slightly lower than the ISM value (\Cratioism{} $= 68 \pm 15$). This suggests potential diversity in \Cratio{} among substellar objects of different ages and spectral types.

    \item Carbon-to-oxygen ratio: The C/O ratio of GQ Lup B ($0.50 \pm 0.01$) is consistent with the lower bound of the solar value. The absence of the higher C/O ratios predicted for formation within a protoplanetary disc supports an origin via gravitational collapse or disc instability. However, interpreting the C/O ratio in the context of planet formation is challenging due to the complex interplay between cloud condensation and atmospheric abundances. This underscores the importance of further studies using isotope ratios as complementary tracers of formation pathways.

    \item Temperature profiles: We introduce a novel retrieval method that uses dynamic pressure-temperature knots to provide flexible temperature profiles and minimise biases in retrieved parameters such as surface gravity. The retrieved temperature profile of GQ Lup B suggests the possible presence of non-adiabatic convection in the lower atmosphere, as evidenced by a reduced temperature gradient.

    \item Impact on isotope ratios: The accurate determination of \Cratio{} depends on precise temperature profiles, as the \thirteenCO{} and \twelveCO{} lines arise from different atmospheric altitudes. This study demonstrates how variations in temperature profiles can lead to differing interpretations of spectral features, which affects retrieved surface gravities, abundances, and isotope ratios.

    \item Veiling in GQ Lup A: Veiling observed in the \textit{K}-band spectrum of GQ Lup A indicates an additional emitting source or photospheric spots on the stellar surface. Veiling was modelled as a wavelength-dependent excess continuum, with the veiling factor fitted independently for each order-detector pair. Observations at longer wavelengths ($>2.5$ \micron) could help determine the physical origin of veiling. Results show that the veiling factor increases with wavelength, consistent with previous measurements by \citealt{sousaNewInsightsNearinfrared2023}.
\end{itemize}

Our results highlight the importance of combining isotopic and elemental ratios to decipher the complex formation histories of substellar objects. Further measurements spanning different ages and spectral types are crucial for advancing our understanding of substellar formation pathways. This study contributes to the broader effort of characterising young planetary and substellar systems by leveraging isotope ratios. We stress the necessity of including isotope measurements of host stars to fully understand the formation history of widely separated companions. Specifically, we present a method for measuring the carbon isotope ratio of a K-type host star, incorporating veiling into the analysis.

The next generation of high-resolution spectrographs will enable the study of substellar companions at smaller separations from their host stars and at fainter magnitudes. The methods outlined in this paper for accounting for starlight contamination can be applied to extract atmospheric properties from such systems. Upcoming mid-infrared spectrographs such as METIS \citep{brandlMETISMidinfraredELT2021} will enable carbon isotope measurements at longer wavelengths. In this regime, understanding veiling and the effects of disc emission on the observed spectrum will be critical for studying young systems.

\begin{acknowledgements}
We thank J. Xuan for providing the temperature profile of GQ Lup B retrieved with KPIC. We thank P. Hauschildt for generating the isotope-dependent PHOENIX models. We are grateful to the anonymous referee for their constructive comments. D.G.P and I.S. acknowledge NWO grant OCENW.M.21.010. Based on observations collected at the European Organisation for Astronomical Research in the Southern Hemisphere under ESO programme(s) 1110.C-4264(F). This work used the Dutch national e-infrastructure with the support of the SURF Cooperative using grant no. EINF-4556. This research has made use of NASA's Astrophysics Data System. This research has made use of adstex (\url{https://github.com/yymao/adstex}).
\newline
\textit{Software}: NumPy \citep{harrisArrayProgrammingNumPy2020}, SciPy \citep{virtanenSciPyFundamentalAlgorithms2020}, Matplotlib \citep{hunterMatplotlib2DGraphics2007}, petitRADTRANS \citep{mollierePetitRADTRANSPythonRadiative2019}, PyAstronomy \citep{czeslaPyAPythonAstronomyrelated2019}, Astropy \citep{collaborationAstropyProjectSustaining2022}, corner \citep{foreman-mackeyCornerPyScatterplot2016}.
\end{acknowledgements}

\bibliography{AutoLibrary}

\onecolumn
\appendix

\section{Observing conditions}\label{app:obs_conditions}
\begin{figure}[h!]
  \centering
  \includegraphics[width=0.5\columnwidth]{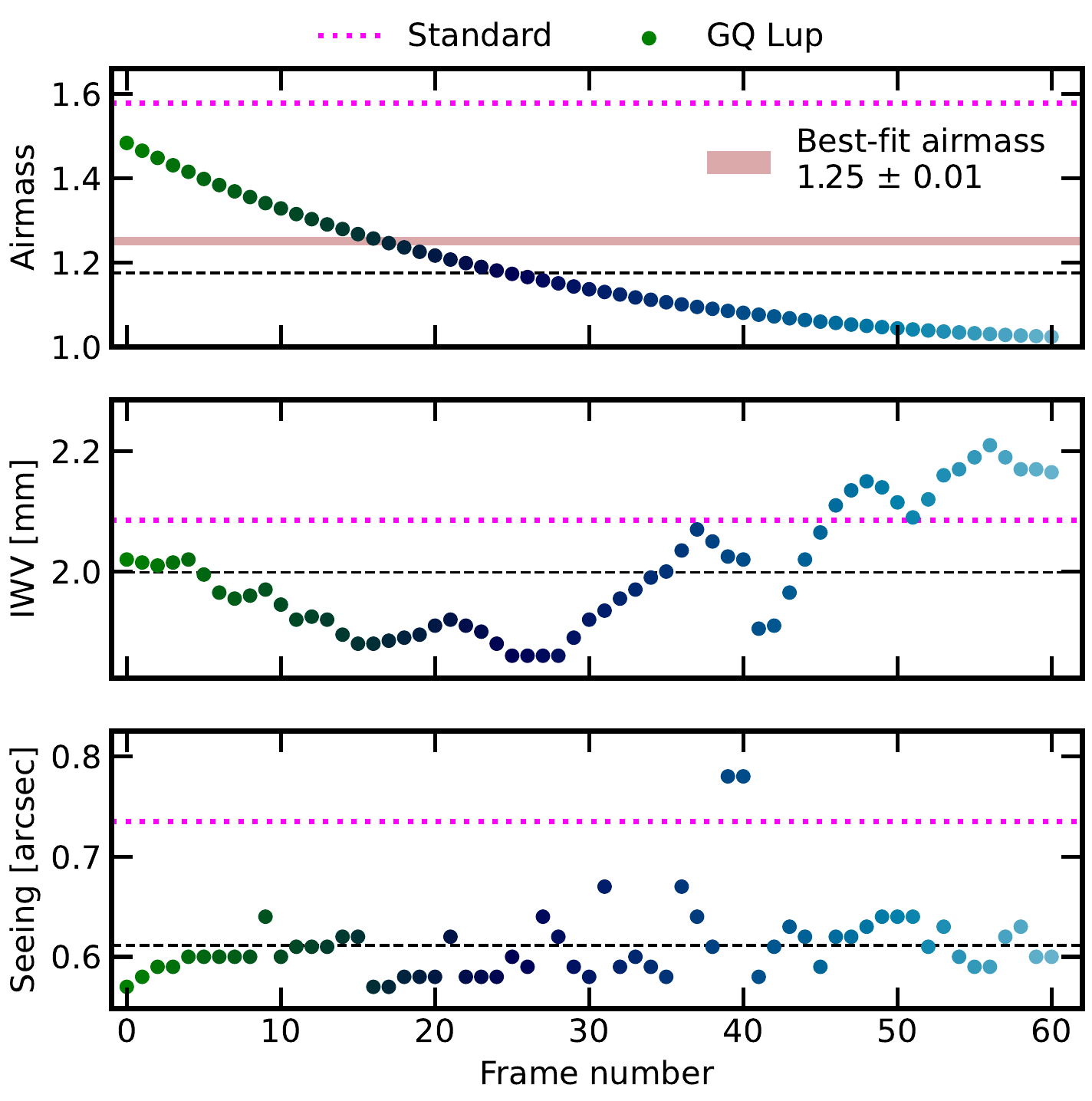}
  \caption{Observing conditions over a 3-hour science exposure sequence. Airmass, integrated water vapour, and seeing are plotted in the top, middle, and bottom panels, respectively. Average values are indicated by dashed black lines. Pink data points represent standard star observations, taken immediately before the science sequence. The retrieved effective airmass, with 1$\sigma$ uncertainty indicated by the line width, is overlaid in the airmass panel.}

  \label{fig:obs_conditions}
\end{figure}

\section{Linear starlight model}\label{app:starlight}
The starlight model is constructed using a linear combination of the observed spectrum of GQ Lup A, where individual components are shifted versions of the on-axis spectrum. In addition, we accounted for low-frequency variations in the scattered starlight using a spline decomposition method, as presented in \citealt{ruffioDetectingExomoonsRadial2023}. The linear model is constructed and fitted for each order-detector pair and nodding position separately, hence providing an efficient method for modelling the starlight spectrum for each chunk of the data without the need to combine data with potential different systematics and noise properties.

\begin{figure}[ht!]
  \centering
  \begin{subfigure}[t]{0.48\textwidth}
    \centering
    \includegraphics[width=\textwidth]{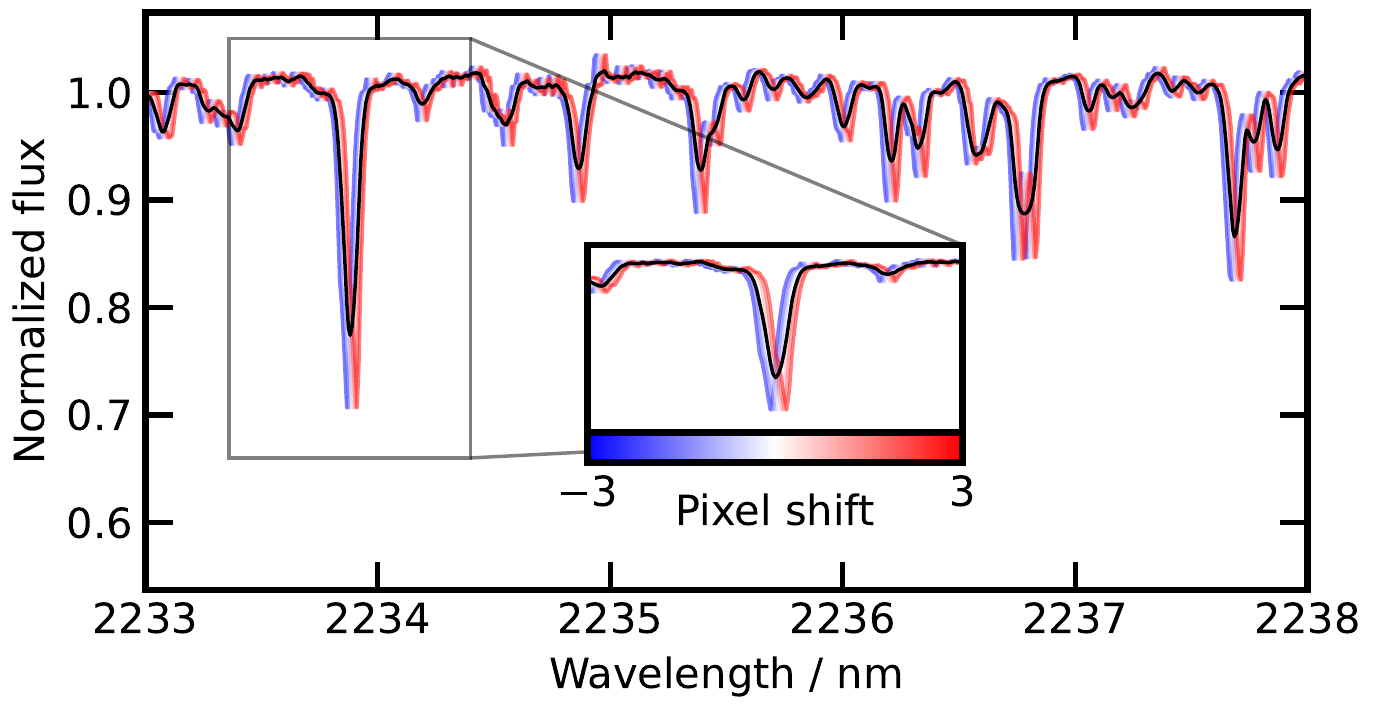}
    \label{fig:starlight_broadening}
  \end{subfigure}
  \hfill
  \begin{subfigure}[t]{0.48\textwidth}
    \centering
      \includegraphics[width=\columnwidth]{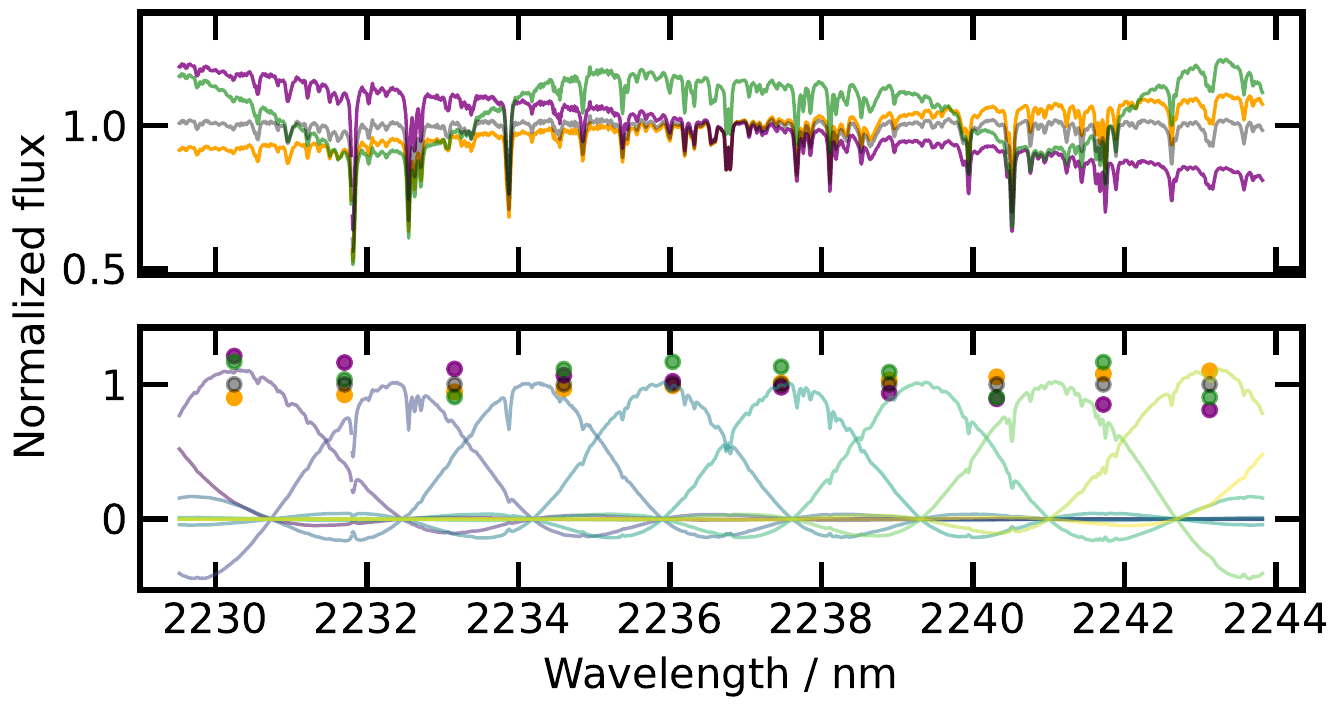}
    \label{fig:starlight_spline}
  \end{subfigure}
  \caption{Components of the starlight model. Left panel: Broadening kernel. The black spectrum represents the combined spectrum with equally weighted components. Right panel: Spline decomposition of the starlight model. The upper part displays the resulting spectra corresponding to different amplitudes, and the lower part the individual spline components. The linear amplitudes used to generate the spectra in the upper part are indicated by markers in the lower part.}
  \label{fig:starlight}
\end{figure}

\clearpage
\section{Extended results of the atmospheric retrieval of GQ Lup B}\label{app:retrieval_gqlup_b}
\begin{figure*}[ht]
  \centering
  \includegraphics[width=0.92\textwidth]{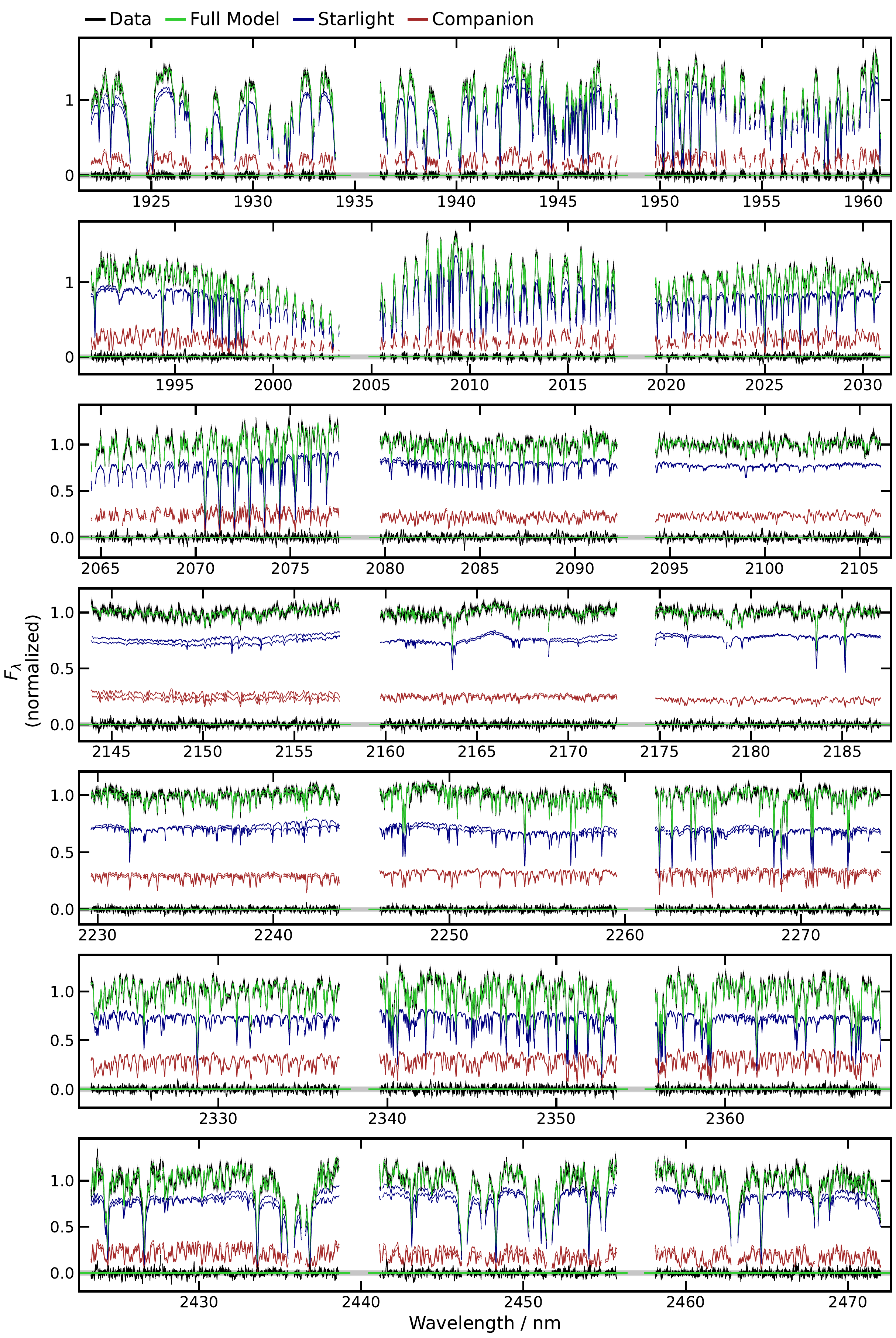}
  \caption{Same as the top panel of \Cref{fig:fig2_bestfit}, but for all spectral orders.}
  \label{fig:bestfit_all_orders}
\end{figure*}

\begin{figure*}[ht!]
  \centering
  \includegraphics[width=\textwidth]{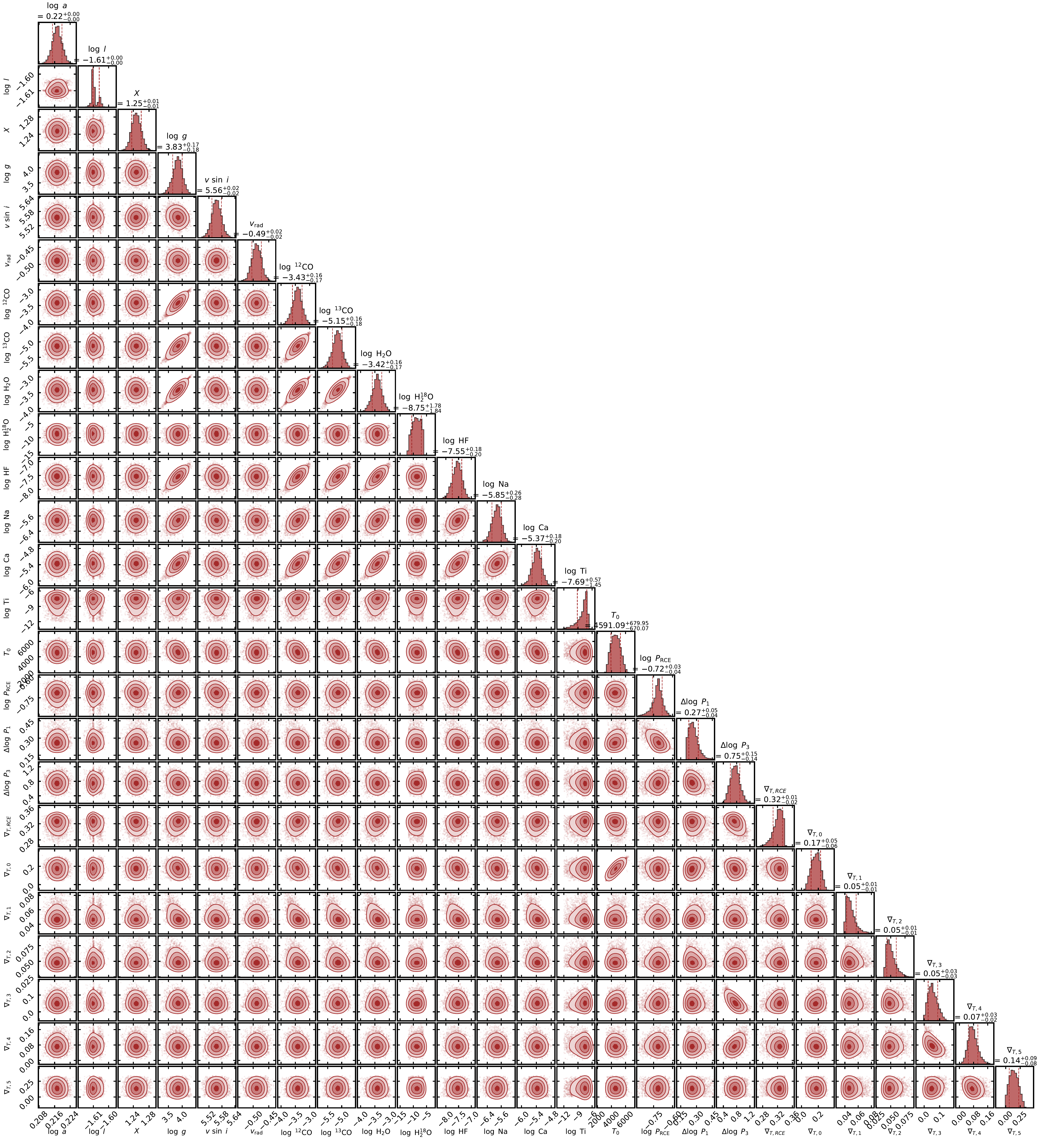}
  \caption{Posterior distributions of the best-fit parameters of GQ Lup B. The uncertainties indicate the 16th and 84th percentiles of the distributions.}
\end{figure*}

\begin{table}[h]
    \caption{Description of the free parameters of the retrieval, prior ranges, and best-fit values.}
    
    \centering
    \renewcommand{\arraystretch}{1.5}
    \begin{tabular}{llll}
    \hline
     Parameter                & Description                                                     & Prior range     & best-fit                 \\
    \hline
     $\log\ \mathrm{^{12}CO}$ & log mixing ratio of \twelveCO                                   & [-12.0, -2.0]   & $-3.4^{+0.2}_{-0.2}$          \\
     $\log\ \mathrm{^{13}CO}$ & log mixing ratio of \thirteenCO                                 & [-12.0, -2.0]   & $-5.2^{+0.2}_{-0.2}$          \\
     $\log\ \mathrm{H_2O}$    & log mixing ratio of $\mathrm{H_2 O}$                            & [-12.0, -2.0]   & $-3.4^{+0.2}_{-0.2}$          \\
     $\log\ \mathrm{HF}$      & log mixing ratio of HF                                          & [-12.0, -2.0]   & $-7.6^{+0.2}_{-0.2}$          \\
     $\log\ \mathrm{Na}$      & log mixing ratio of Na                                          & [-12.0, -2.0]   & $-5.8^{+0.3}_{-0.3}$          \\
     $\log\ \mathrm{Ca}$      & log mixing ratio of Ca                                          & [-12.0, -2.0]   & $-5.4^{+0.2}_{-0.2}$          \\
     $\log\ \mathrm{Ti}$      & log mixing ratio of Ti                                          & [-12.0, -2.0]   & $-7.7^{+0.6}_{-1.4}$          \\
     $\log\ g$ [cm/s$^2$]              & log surface gravity                                             & [2.0, 5.0]      & $3.83^{+0.17}_{-0.18}$        \\
     $v\ \sin\ i$ [km/s]            & projected rotational velocity                                   & [2.00, 20.00]   & $5.56^{+0.02}_{-0.02}$        \\
     $v_\mathrm{rad}$ [km/s]        & radial velocity                                                 & [-20.00, 20.00] & $-0.49^{+0.02}_{-0.02}$       \\
     $\nabla_{T,0}$           & temperature gradient at $P_0=10^2$ bar                          & [0.04, 0.34]    & $0.17^{+0.05}_{-0.06}$        \\
     $\nabla_{T,1}$           & temperature gradient at $P_\mathrm{RCE}+2\Delta P_\mathrm{bot}$ & [0.04, 0.34]    & $<0.06$        \\
     $\nabla_{T,2}$           & temperature gradient at $P_\mathrm{RCE}+\Delta P_\mathrm{bot}$  & [0.04, 0.34]    & $<0.06$        \\
     $\nabla_{T,\mathrm{RCE}}$         & temperature gradient at $P_\mathrm{RCE}$                        & [0.04, 0.34]    & $0.32^{+0.01}_{-0.02}$        \\
     $\nabla_{T,3}$           & temperature gradient at $P_\mathrm{RCE}-\Delta P_\mathrm{top}$  & [0.00, 0.34]    & $0.05^{+0.03}_{-0.03}$        \\
     $\nabla_{T,4}$           & temperature gradient at $P_\mathrm{RCE}-2\Delta P_\mathrm{top}$ & [0.00, 0.34]    & $0.07^{+0.03}_{-0.02}$        \\
     $\nabla_{T,5}$           & temperature gradient at $P_5=10^{-5}$ bar                       & [0.00, 0.34]    & $0.14^{+0.09}_{-0.08}$        \\
     $\Delta\log\ P_\mathrm{bot}$ [bar]        & log pressure shift of lower PT knots                            & [0.20, 1.60]    & $0.27^{+0.05}_{-0.04}$        \\
     $\Delta\log\ P_\mathrm{top}$ [bar]       & log pressure shift of upper PT knots                            & [0.20, 1.60]    & $0.75^{+0.15}_{-0.14}$        \\
     $\log\ P_\mathrm{RCE}$ [bar]  & log pressure of RCE                                             & [-3.00, 1.00]   & $-0.72^{+0.03}_{-0.04}$       \\
     $T_0$ [K]               & temperature at $10^{2}$ bar                                     & [2000, 10000]   & $4591^{+680}_{-670}$          \\
     $X$                      & airmass                                                         & [0.70, 1.60]    & $1.25^{+0.01}_{-0.01}$        \\
     $\log\ a$                & GP amplitude                                                    & [-2.00, 0.40]   & $0.2173^{+0.0025}_{-0.0024}$  \\
     $\log\ l$ [nm]               & GP length scale                                                  & [-2.00, -1.20]  & $-1.6100^{+0.0038}_{-0.0003}$ \\
    \hline
    \end{tabular}
    \label{tab:free_params}
    \parbox{0.9\linewidth}{\small
\textit{Notes:} The best-fit values are the median values of the posterior distribution, with the 1$\sigma$ uncertainty given by the 16th and 84th percentiles. The best-fit values with 1$\sigma$ lower limits are indicated by $<$ when the posterior distribution is truncated at the edge of the prior range.}
\end{table}
\clearpage

\begin{figure}[h]
    \centering
    \includegraphics[width=\columnwidth]{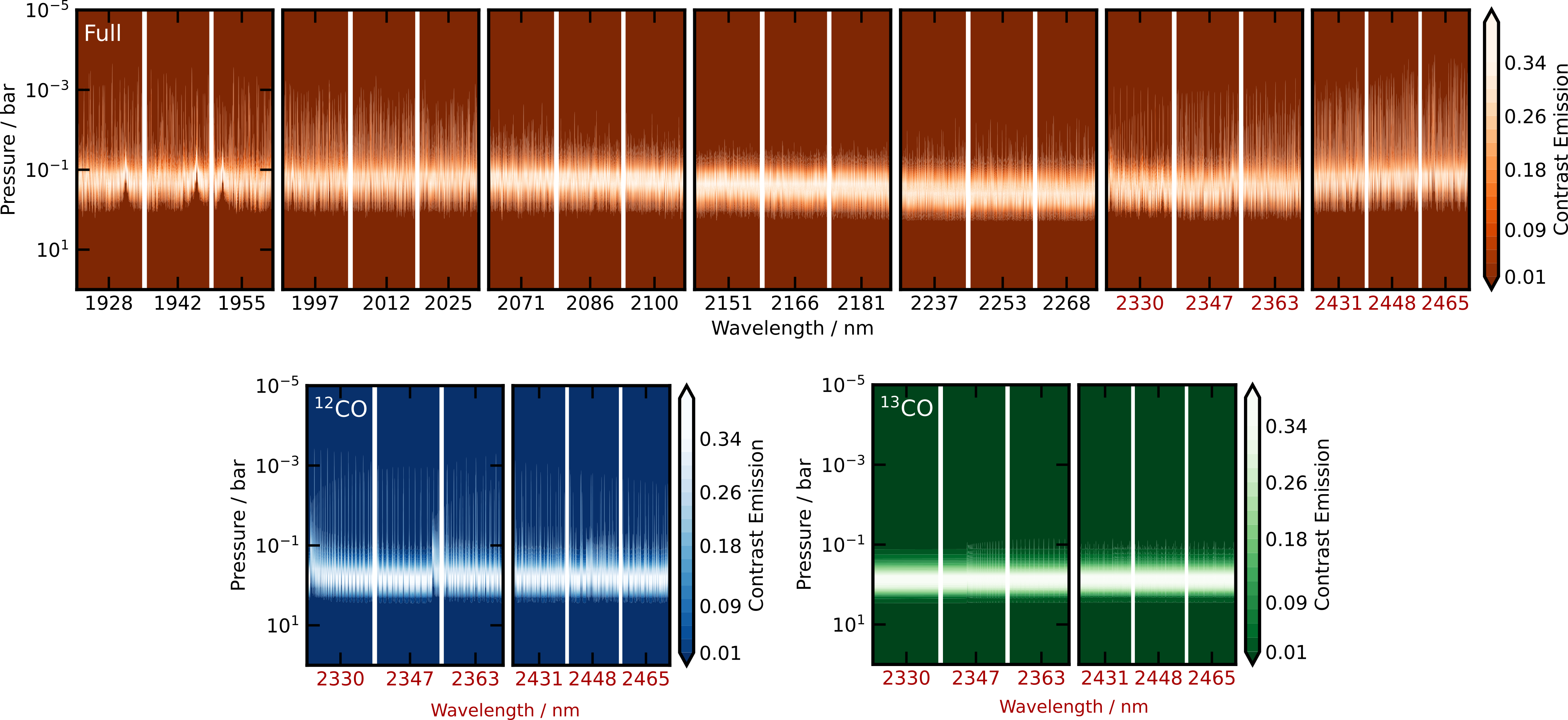}
    \caption{Contribution function of the full best-fit model of GQ Lup B (top panel). The individual contribution functions of \twelveCO{} and \thirteenCO{} are displayed for the last two spectral orders on the bottom panels.}
    \label{fig:contr_em_CO}
\end{figure}

\section{Veiling in the spectrum of GQ Lup A}\label{app:veiling_gqlup_a}
\begin{figure*}[ht!]
  \centering
  \includegraphics[width=\textwidth]{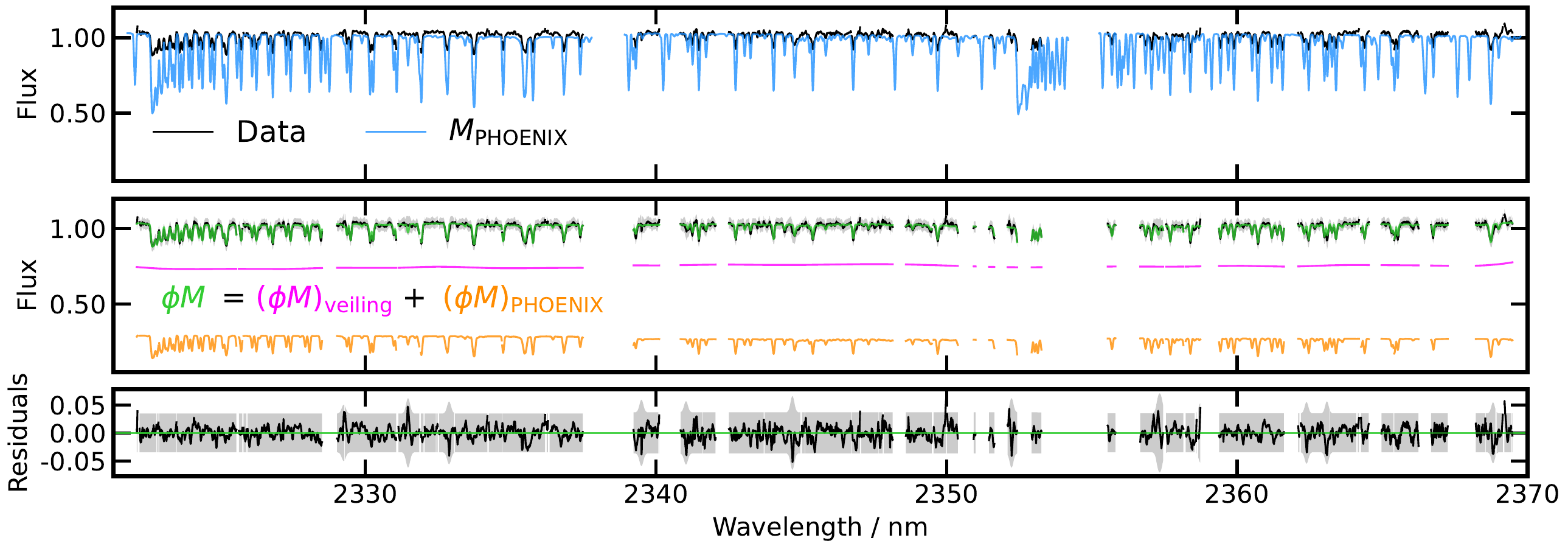}
  \caption{Observed spectrum of GQ Lup A (black) for the spectral order centred at 2345 nm, compared with a PHOENIX template ($T_{\text{eff}} = 4300$ K, $\log{g} = 4.0$) that includes instrumental and rotational broadening. The mismatch between the observed line depths and the model without veiling is evident. The second panel shows the best-fit linear model (green), which incorporates a veiling continuum (pink) alongside the PHOENIX model (orange). Residuals of the best-fit model are presented in the third panel.}
  \label{fig:gqlupa_bestfit_linear_model}
\end{figure*}
\newpage
\section{Posterior comparison with fixed temperature profile from X24}
\begin{figure}[h]
    \centering
    \includegraphics[width=0.97\columnwidth]{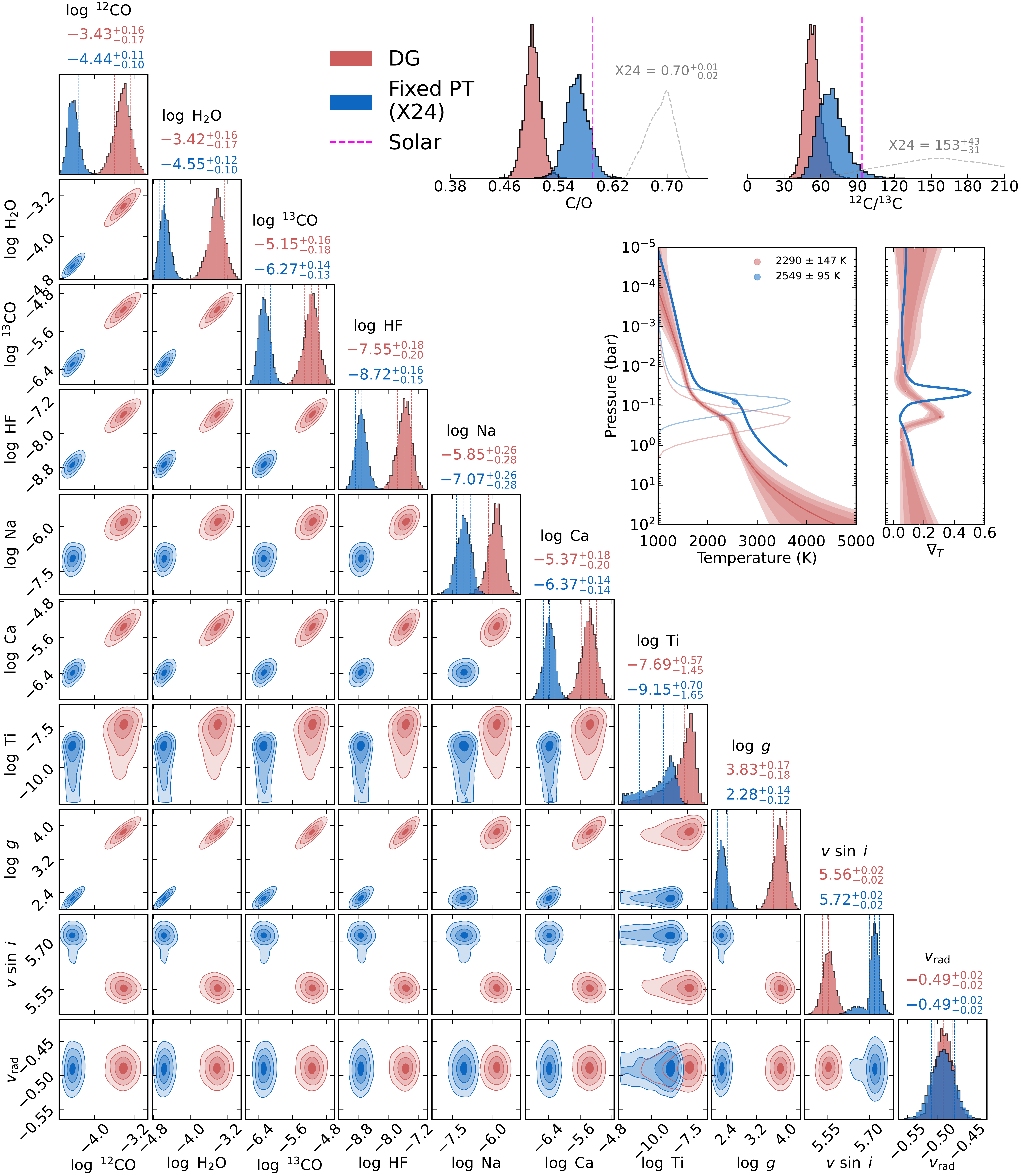}
    \caption{Comparison of posterior distributions for selected best-fit parameters of GQ Lup B, highlighting differences between our fiducial model (with DG) and the fixed PT profile model from \citealt{xuanAreThesePlanets2024} (X24). Dashed lines in the corner plot represent the 16th and 84th percentiles of the distributions. Derived C/O and \Cratio{} values are displayed in the top-right corner, with the dashed grey probability density indicating the values reported by X24. The middle-right panel shows the retrieved temperature profiles, while the rightmost panel presents the corresponding temperature gradients.}
    \label{fig:corner_fixed_PT}
\end{figure}

\end{document}